\documentstyle[12pt]{article}
 \let\msk=\medskip 
\let\nn=\nonumber
\let\th=\vartheta
\def\Th{\4\theta}

\let\qd=\quad  
\let\ep=\varepsilon

\def\up{{\underline p}} \def\op{{\overline p}}

\def\s0#1#2{\mbox{\small{$\frac{#1}{#2}$}}}
\def\5{\bar }  \def\6{\partial } \def\7{\hat } \def\4{\tilde }
\let\LRA=\Leftrightarrow
\let\then=\Rightarrow

\def\bea{\begin{eqnarray}} \def\eea{\end{eqnarray}}
\def\beann{\begin{eqnarray*}} \def\eeann{\end{eqnarray*}}
\def\beq{\begin{equation}} \def\eeq{\end{equation}}
\def\ba{\begin{array}} \def\ea{\end{array}}
\def\ben{\begin{enumerate}} \def\een{\end{enumerate}}

 \def\cB{{\cal B}} \def\cG{{\cal G}}
\def\cD{{\cal D}}  \def\cA{{\cal A}}
\def\cR{{\cal R}}  
\def\cF{{\cal F}} \def\cT{{\cal T}{}} \def\cH{{\cal H}}
\def\cW{{\cal W}} \def\cP{{\cal P}} \def\cS{{\cal S}}
 \def\cU{{\cal U}} \def\cV{{\cal V}}

\def\TIX{r}
\def\da{{\dot \alpha}} \def\dbe{{\dot \beta}}
\def\dg{{\dot \gamma}} 
\def\ua{{\underline{\alpha}}} \def\ube{{\underline{\beta}}}

\def\dR{{(r)}}
\def\T#1#2#3{{T_{#1#2}}^{#3}} 
\def\F#1#2#3{{F_{#1#2}}^{#3}}
\def\G#1#2#3{{G_{#1#2}}^{#3}}
\def\Gg#1#2#3{{g_{#1#2}}^{#3}}
\def\f#1#2#3{{f_{#1#2}}^{#3}}
\def\A#1#2{{A_{#2}}^{#1}}
\def\spin#1#2{\omega_{#2}{}^{#1}}
\def\RS #1#2{\psi_{#2}{}^{#1}}
\def\bRS #1#2{\5\psi_{#2}{}^{#1}}
\def\viel#1#2{e_{#2}{}^{#1}}
\def\Viel#1#2{E_{#1}{}^{#2}}

\def\csum#1#2{\sum_{#1}\hspace{-1.#2em}\circ\ \ \ }

\newcommand{\mysection}[1]{\section{#1}
            \setcounter{equation}{0}\setcounter{figure}{0}}

\def\gh{\mbox{gh}} 
\def\tot{\mbox{totdeg}}
\def\deg{\mbox{formdeg}}
\def\tdeg#1{(#1)}

\begin{document}
\begin{titlepage}

\begin{flushright}

Ann.\ Phys.\ (N.Y.)\ {\bf 259} (1997) 253--312\\
hep-th/9609192\\

\end{flushright}
\vfill

\begin{center}
{\Huge  Local BRST Cohomology in Minimal\\[.5ex]
D=4, N=1 Supergravity\\[.5ex]}
\end{center}
\vfill

 \begin{center}
 {\Large
 Friedemann Brandt}
 \end{center}
 \vfill

 \begin{center}{\sl
 Instituut voor Theoretische Fysica, Katholieke
 Universiteit Leuven,\\
 Celestijnenlaan 200 D, B--3001 Leuven, Belgium\\
and\\
 Departament d'Estructura i Constituents de la Mat\`eria,
 Facultat de F\'{\i}sica,
 Universitat de Barcelona,
 Diagonal 647,
 E-08028 Barcelona, Spain.\,$^*$
 }\end{center}
 \vfill

\begin{abstract}
The local BRST cohomology is computed in old and new minimal supergravity, 
including the coupling to Yang-Mills gauge multiplets. This covers the 
determination of all gauge invariant local actions for these models, the 
classification of all the possible counterterms that are invariant on-shell, 
of all candidate gauge anomalies, and of the possible nontrivial (continuous)
deformations of the standard actions and gauge transformations. Among
others it is proved that in old minimal supergravity the most general 
gauge invariant action can indeed be constructed from well-known superspace 
integrals, whereas in new minimal supergravity there are only a few 
additional (but important) contributions which cannot be constructed in 
this way without further ado. Furthermore the results indicate
that supersymmetry itself is not anomalous in minimal supergravity and show 
that the gauge transformations are extremely stable under consistent 
deformations of the models. There is however an unusual deformation 
converting new into old minimal supergravity with local $R$-invariance 
which is reminiscent of a duality transformation.
\end{abstract}

\vfill

\hrule width 5.cm
\vspace*{.5em}
{\small \noindent $^*$ Present address. E-mail: brandt@ecm.ub.es.}

\end{titlepage}


\mysection{Introduction}\label{intro}

\subsection{Topics and motivation}

This paper presents the first exhaustive analysis of the local BRST 
cohomology in supergravity. More precisely we will analyze the two 
most popular formulations of four dimensional $N=1$ supergravity 
known as old \cite{old} and new \cite{new} minimal
supergravity, including their coupling to
Yang--Mills gauge multiplets. The inclusion of further (matter)
multiplets will be briefly discussed too. 

To begin with, let me recall some general features of the local 
BRST cohomology. By this I mean the cohomology of the BRST operator%
\footnote{Throughout the paper, the word ``BRST operator"
describes the strictly nilpotent antiderivation which acts
nontrivially on both the fields and the antifields and is
generated in the antibracket by the solution of the master
equation \cite{bv}, see section \ref{brs} for details.} 
in the space of local functionals of the fields and antifields.
It allows to analyze various physically relevant aspects
of gauge theories in a unified framework.
Historically it was above all the anomaly problem that
initiated the interest in this cohomology when it became
clear that the possible gauge anomalies define cohomology
classes at ghost number one. This was first shown
in abelian Higgs--Kibble and Yang--Mills theories \cite{brs1,brs} 
and later generalized
to arbitrary gauge theories \cite{TvNvP}. 

The classification of possible anomalies is however only one
instance that can be efficiently investigated by means of the
local BRST cohomology. Further well-known applications of this
cohomology are the construction of gauge invariant actions
and the classification of the possible
``on-shell counterterms". Both the actions and these
counterterms are local functionals of the classical fields, 
the difference being that actions have to be
gauge invariant off-shell whereas the counterterms
need to be gauge invariant only on-shell. 
Gauge invariant actions and on-shell counterterms define 
BRST cohomology classes at ghost number zero.

Possibly still less known are other applications of the
local BRST cohomology which were realized more recently and
will be therefore briefly described in a little 
more detail in the following.

One of these applications concerns 
the problem whether or not a given gauge
theory, defined through a particular gauge invariant action,
can be consistently and continuously deformed in a nontrivial
manner. Here a deformation is called {\em consistent}
when the deformed action is still gauge invariant
but under possibly modified gauge transformations,
{\em continuous} when it can be parametrized by a deformation parameter
$g$ such that one recovers the original theory for $g=0$, and
{\em nontrivial} when the deformation cannot be removed through
mere field redefinitions. Such deformations concern
thus simultaneously the action {\em and} the gauge transformations.
To first order in $g$ they are determined by the local
BRST cohomology at ghost number zero \cite{bh}. At higher order
in $g$ they might get additionally obstructed by the local BRST 
cohomology at ghost number one \cite{brandt}.

Last but not least the local BRST cohomology provides, at
{\em negative} ghost numbers, the 
dynamical conservation laws of a theory \cite{bbh1}. These
conservation laws are
described in terms of totally antisymmetric local functions 
$j^{\mu_1\dots\mu_k}$
of the classical fields with on-shell vanishing divergence
($\6_{\mu_1}j^{\mu_1\dots\mu_k}\approx 0$), defined
modulo trivial
conservation laws which are on-shell of the form 
$\6_{\mu_0}k^{[\mu_0\dots\mu_k]}$. As shown recently 
in \cite{bhw}, {\em each} dynamical conservation law
corresponds to a global (= rigid) symmetry of 
the solution of the master equation \cite{zj,bv} and gives rise
to a corresponding Ward identity. This generalizes Noether's
theorem on the correspondence of dynamically conserved
currents (= conservation laws of 
order $k=1$) and the global symmetries of the 
{\em classical} action. 

To summarize, the local BRST cohomology contains physically
relevant information 
at {\em all} ghost numbers $\leq 1$ (a physical interpretation
of the cohomology classes at ghost numbers $>1$ has not yet been
found). Therefore it is worthwhile to analyze this cohomology 
at least for these ghost numbers.

The analysis of the local BRST cohomology
in minimal supergravity carried out here will allow us 
to answer questions which in part have been discussed already
in the early days of supergravity without having been
answered exhaustively so far. For instance, the problem whether the
local supersymmetry transformations can be modified
nontrivially was raised already in \cite{sugradefs} soon after 
supergravity was invented in \cite{invention}. 
Shortly after that, the discussion of the
possible on-shell counterterms was started \cite{PvN}, mainly
to clarify whether or up to which loop order
supergravity may be finite in the conventional
quantum field theoretical approach. The construction
of supergravity actions has been discussed extensively in the
literature because it is much more
involved than in standard gravity. In particular various
superspace techniques have been invented for this purpose,
see e.g. the textbooks \cite{wessbagger,bible}. In this paper
we will clarify whether or to what extend
such methods provide the {\em most general} local
action functional for old and new minimal supergravity.
Anomalies in supergravity have been already discussed 
in the literature using the BRST approach,
see e.g. \cite{bonora}. However, the work on this topic
concentrated up to now mainly on the 
determination of chiral anomalies in supergravity, 
leaving open among others the important question 
whether or not local supersymmetry itself can be anomalous.

\subsection{Sketch of the approach}

To compute the local BRST cohomology for old and new minimal 
supergravity, one has to solve
\beq s\omega_4+d\omega_3=0\label{i1}\eeq
where $\omega_4$ is the integrand of a local functional, written
as a differential 4-form, $\omega_3$ is some local 3-form, $s$ is the
nilpotent BRST operator for the supergravity theories studied here,
and $d$ denotes the spacetime exterior derivative. The searched for
solutions $\omega_4$ are defined
modulo trivial solutions of the form $s\eta_4+d\eta_3$. In particular
$\omega_4$ itself is of course required to be nontrivial,
\beq \omega_4\neq s\eta_4+d\eta_3.\label{i1b}\eeq

We will investigate (\ref{i1}) and (\ref{i1b}) for all ghost numbers
but spell out the results in detail only for the physically most
important cases, i.e.\ for ghost numbers $\leq 1$.
The investigation will be truly general, i.e.\ it will not
use restrictive assumptions on the form of the solutions.
For instance, a restriction on the
number (and order) of derivatives of the fields and antifields occurring 
in $\omega_4$, $\omega_3$, $\eta_4$ and $\eta_3$
will not be imposed, apart from requiring 
this number to be finite%
\footnote{More precisely we require finiteness of the number
of derivatives in every term occurring in an expansion of 
the forms according to the number of antifields. This defines ``local
forms" here.}.
Furthermore, the solutions will not be assumed
to be covariant in whatever sense from the outset. In particular it will
neither be assumed that they transform as true differential forms under
spacetime diffeomorphisms, nor that any
group indices of the fields occurring in them
are ``correctly" contracted. Also, we will of course not assume
that the solutions can be constructed
in whatsoever fashion from superfields
(superfields and superspace techniques
will nowhere be used in this paper!). 
The cohomological analysis itself will therefore reveal
to what extend the solutions have such properties.

Our only inputs will be the field content and the BRST transformations.
We will use the
standard set of auxiliary fields to close the gauge algebra
off-shell and the standard actions and gauge transformations
to construct the BRST operator. These gauge transformations 
are of course general coordinate
transformations (= spacetime diffeomorphisms),
$N=1$ supersymmetry transformations, Lorentz and
Yang--Mills transformations, as well as
the reducible gauge transformations of
the 2-form gauge potential present in
new minimal supergravity. 

The use of the auxiliary fields is
in principle not necessary as the antifield formalism 
allows to treat also open gauge algebras \cite{bv}. Nevertheless it has
some advantages to close the gauge algebra off-shell by means
of auxiliary fields. For instance this will facilitate to
distinguish nontrivial deformations of the
gauge transformations from trivial ones, and counterterms
which are invariant only on-shell from those which can even be
completed to off-shell invariants by means of the auxiliary fields.
As a matter of fact, the solutions of (\ref{i1}) and (\ref{i1b}) 
in a formulation without
auxiliary fields can be easily obtained from those given here.
To that end one must eliminate the auxiliary fields using
their `generalized equations of motion'
resulting from the solution of the master equation rather than
from the classical action, see \cite{auxfields,noetheranos} for details.

To analyze (\ref{i1}) and (\ref{i1b}) systematically we will use a 
general framework described in \cite{ten} (see also \cite{tenold}).  
This reduces the
computation of the BRST cohomology (locally) to a 
particular ``covariant" cohomological problem 
for the operator $\4s=s+d$ in the space of {\em local total forms}
depending on suitably defined {\em tensor fields} and 
{\em generalized connections}. Local total forms are simply formal
sums of local differential forms 
with different form degrees and
ghost numbers. 
To solve the reduced problem we will then combine results and
techniques (e.g.\ Lie algebra cohomology) used successfully already
in nonsupersymmetric theories \cite{com,grav,bhprl,bbh2,bbhgrav}
and methods developed in \cite{phd,glusy} to deal with supersymmetry.

\subsection{Outline of the paper}

We will first define the problem precisely by fixing
the field content and the BRST transformations in
section \ref{brs}. There we will also introduce already
the relevant generalized connections and 
tensor fields. It will be crucial for the analysis to
determine an appropriate complete set of
these tensor fields such that there are no 
identities (including ``differential" ones) 
relating the elements of this set off-shell. This will be done
in section \ref{basis}.

We will then be prepared to compute the local BRST cohomology.
This computation will be carried out in two separate steps. 
In section \ref{rest} we will first compute
the {\em restricted BRST cohomology} defined through
(\ref{i1}) and (\ref{i1b}) in the space of antifield independent
local forms. Thanks to the use of the auxiliary fields,
this problem is well-defined and reduces locally to the corresponding  
$\4s$-cohomology in the space of antifield independent total
local forms depending only on tensor fields and generalized connections.
Its solution will provide in particular the most general action
functionals for old and new minimal supergravity spelled out in
section \ref{actions}. 

In section \ref{full} we will
then analyze the {\em full BRST cohomology}, i.e. the
solutions of (\ref{i1}) and (\ref{i1b}) in the space of local forms which
may depend on antifields as well. This problem can (and will) be also 
traced back to a cohomological problem defined in the space
of total local forms depending only on the tensor fields and generalized 
connections (but not on antifields); however, this time the relevant 
cohomology is the {\em weak (= on-shell) $\4s$-cohomology} 
in this space \cite{ten}. A comparision of the results for the
restricted and the weak $\4s$-cohomology will finally enable us to
distinguish those solutions which necessarily involve antifields
from those which can be written entirely in terms of the fields when the
auxiliary fields are used (note however that
the elimination of the auxiliary fields introduces in general additional
antifield dependence \cite{auxfields}).
The results for the local conservation laws, on-shell counterterms,
consistent deformations and for the candidate gauge anomalies
are presented in sections 
\ref{noether}, \ref{counter} and \ref{anos}.

In section \ref{matter} we discuss the modifications
of the results when further (matter) multiplets are included
or a more complicated action is used from the start. Section \ref{top}
briefly comments on topological aspects which are
neglected in the remainder of the paper.
A summary of the main results with some concluding comments is
given in section \ref{conclusion}, followed by
several appendices containing conventions used in
this paper, details concerning the realization of supersymmetry
in minimal supergravity which underlie crucially the results,
and the derivation of two theorems used within the computation.
\newpage

\mysection{Field content and BRST algebra}
\label{brs}

Old and new minimal supergravity differ both in their field content
and gauge symmetries. The field content 
of old minimal supergravity in a formulation with closed gauge algebra,
including Yang--Mills gauge multiplets and all the ghosts, is summarized
in table 2.1 where $\gh(\Phi)$, $\ep(\Phi)$ and $dim(\Phi)$ denote the
ghost number, Grassmann parity and dimension of $\Phi$ respectively
(the dimension assignments are the `natural' ones).
\[
\ba{c|c|c|c|l}
\Phi             &  \gh(\Phi) & \ep(\Phi) & dim(\Phi) & \\
\hline\rule{0em}{3ex}
\viel a\mu       &       0    &     0     &  0   &
                \mbox{vierbein}\\
\psi_\mu         &       0    &     1     &  1/2 &
                \mbox{gravitino} \\
M                &       0    &     0     &   1  &
                \mbox{complex Lorentz scalar (aux.)}\\
B_a              &       0    &     0     &   1  &
                \mbox{real Lorentz vector (aux.)}\\
C^\mu     &       1    &     1     &  -1  &
                \mbox{diffeomorphism ghosts}\\
\xi              &       1    &     0     & -1/2 &
                \mbox{supersymmetry ghosts}\\
C^{ab}           &       1    &     1     &  0   &
                \mbox{Lorentz ghosts}\\
\hline\rule{0em}{3ex}
\A i\mu          &       0    &     0     &   1  &
                \mbox{Yang-Mills gauge fields}\\
\lambda^i        &       0    &     1     &  3/2 &
                \mbox{gauginos}\\
D^i              &       0    &     0     &   2  &
                \mbox{real Lorentz scalars (aux.)}\\
C^i              &       1    &     1     &   0  &
                \mbox{Yang-Mills ghosts}\\
\multicolumn{5}{c}{}\\
\multicolumn{5}{c}{\mbox{Table 2.1:
Field content of old minimal supergravity}}
\ea
\]
The new minimal supergravity
multiplet contains instead of the auxiliary fields
$M$ and $B_a$ a 2-form gauge
potential and a gauge field
for $R$-transformations, with corresponding ghosts and
a ghost for ghosts, cf.\ table 2.2.
Accordingly the gauge symmetries of new minimal supergravity 
include the reducible gauge transformations of 
the 2-form gauge potential and local $R$-invariance.
In contrast, old minimal supergravity may or may not be locally
$R$-invariant (both cases are covered by our analysis).
The Yang--Mills gauge multiplets have in new minimal supergravity 
the same field content ($\A i\mu,\lambda^i,D^i$) as in old minimal 
supergravity, except for the missing
gaugino and $D$-field for $R$-transformations (when old minimal
supergravity with local $R$-symmetry is considered, the
corresponding gauge field, gaugino and $D$-field
count among the $\A i\mu,\lambda^i,D^i$ in table 2.1).
\[
\ba{c|c|c|c|l}
\Phi             &  \gh(\Phi) & \ep(\Phi) & dim(\Phi) & \\
\hline\rule{0em}{3ex}
t_{\mu\nu}       &       0    &     0     &    0      &
    \mbox{2-form gauge potential}\\
Q_\mu            &       1    &     1     &    -1     &
    \mbox{ghosts associated with $t_{\mu\nu}$} \\
Q                &       2    &     0     &    -2     &
    \mbox{ghost for the ghosts $Q_\mu$}\\
\hline\rule{0em}{3ex}
\A \dR\mu        &       0    &     0     &    1      &
    \mbox{gauge field for $R$-transformations}\\
C^\dR            &       1    &     1     &    0      &
    \mbox{ghost for $R$-transformations}\\
\multicolumn{5}{c}{}\\
\multicolumn{5}{c}{\mbox{Table 2.2: Fields in
new minimal supergravity replacing $M$ and $B_a$}}
\ea
\]

In order to define the cohomological problem we need to
specify the
BRST transformations of all the fields in tables 2.1 and 2.2 and of
their antifields. Since a formulation
with a closed gauge algebra is used,
the BRST transformations of the `classical fields' (i.e.\ those
fields with ghost number 0) can be obtained directly from
their gauge transformations by replacing the gauge parameters 
with the corresponding ghosts, using the standard field variations under
diffeomorphisms, Lorentz and Yang--Mills transformations, 
the supersymmetry transformations given  e.g.\ in 
\cite{old,new,wessbagger}, as well as the extra gauge
transformations in new minimal supergravity. 
For instance the BRST transformation
of the vierbein fields reads
\beq
s\viel a\mu = C^\nu\6_\nu\viel a\mu+(\6_\mu C^\nu)\viel a\nu
+C_b{}^a\viel b\mu+2i(\xi\sigma^a\5\psi_\mu-\psi_\mu\sigma^a\5\xi).
\label{sviel}\eeq
The BRST transformation of the ghosts (and the ghost for ghosts)
are then chosen such that the BRST operator is nilpotent. In old minimal
supergravity this yields
\bea
sC^\mu&=&C^\nu\6_\nu C^\mu+2i\,\xi\sigma^\mu\5\xi,
\label{sCdiff}\\
s\xi^\alpha&=&C^\mu\6_\mu \xi^\alpha
+\s0 12C^{ab}{\sigma_{ab\,\beta}}^\alpha\xi^\beta
+i\, C^\dR\xi^\alpha
-2i\, \xi\sigma^\mu\5\xi\,\RS \alpha\mu,\label{sxi}\\
sC^i&=&C^\mu\6_\mu C^i +\s0 12 \f jki C^k C^j
-2i\, \xi\sigma^\mu\5\xi\,\A i\mu,
\label{sCym}\\
sC^{ab}&=&C^\mu\6_\mu C^{ab}
+C^{ca}{C_c}^b-2i\, \xi\sigma^\mu\5\xi\,\spin {ab}\mu\\
& &+\s0 12\,\xi\sigma^{ab}\xi \, \5M
+\s0 12\,\5\xi\5\sigma^{ab}\5\xi \, M
+2i\,\ep^{abcd}\xi\sigma_c\5\xi \, B_d
\label{sC}\eea
where $\f jki$ are the structure constants of the Lie algebra
of the Yang--Mills gauge group and
$\spin {ab}\mu$ is the usual gravitino dependent
spin connection, cf.\ appendix \ref{appA}.

The BRST transformations obtained in this way for 
old minimal supergravity with local $R$-symmetry turn into 
those for new minimal supergravity
by setting $M$ to zero (off-shell) and identifying $B_a$ with
\beq B^a\equiv\s0 16\,\ep^{abcd}H_{bcd},\quad
H_{\mu\nu\rho}=3\6_{[\mu}t_{\nu\rho]}
+6i\psi_{[\mu}\sigma_\nu\5\psi_{\rho]}\ .
\label{identify}\eeq
Furthermore one gets the
following BRST transformations of $t_{\mu\nu}$, $Q_\mu$ and
$Q$:
\bea
st_{\mu\nu}&=&\6_\nu Q_\mu-\6_\mu Q_\nu
+C^\rho\6_\rho t_{\mu\nu}+(\6_\mu C^\rho)\, t_{\rho\nu}
+(\6_\nu C^\rho)\, t_{\mu\rho}\nonumber\\
& &-i\,(\xi\sigma_\mu\5\psi_\nu-\xi\sigma_\nu\5\psi_\mu
+\psi_\mu\sigma_\nu\5\xi-\psi_\nu\sigma_\mu\5\xi),
\label{st}\\
sQ_\mu&=&\6_\mu Q+C^\nu\6_\nu Q_\mu+(\6_\mu C^\nu)Q_\nu
-2i\, \xi\sigma^\nu\5\xi\, t_{\mu\nu}-i\, \xi\sigma_\mu\5\xi,
\label{sQmu}\\
sQ&=&C^\mu\6_\mu Q-2i\, \xi\sigma^\mu\5\xi \, Q_\mu\ .
\label{sQ}\eea

Finally, the BRST transformation
of the antifields are obtained
from the respective solution $\cS$ of the (classical) master equation
\cite{bv} through
\beq s\Phi^*_A=\frac{\delta^R \cS}{\delta \Phi^A}\ .
\label{ss0}\eeq
Thanks to the closure of the gauge algebra, $\cS$
itself is simply given by
\beq
\cS=\cS_{cl}-\int d^4x\, (s\Phi^A)\Phi^*_A
\label{s1}\eeq
where $\cS_{cl}$ denotes the classical action and
$\{\Phi^A\}$ is the
set of fields given in tables 2.1 and 2.2.%
\footnote{Since antighosts and the corresponding
Nakanishi-Lautrup (Lagrange multiplier)
fields and their antifields contribute only via trivial
solutions to the cohomological problem (see e.g. \cite{com,bbh1}),
they are completely neglected in this paper
without loss of generality.}

The somewhat tedious construction of the BRST algebra outlined 
above can in fact be streamlined considerably. Namely it
can be obtained more easily and elegantly
from the gauge covariant supergravity algebra which is well-known
in the  literature and summarized in appendix \ref{appA}.
In particular this provides the BRST algebra
directly in a compact form which is best suited for
the cohomological analysis \cite{ten}.
The gauge covariant algebra reads
\beq
[\cD_A,\cD_B\}=-\T ABC\cD_C-\F ABI\delta_I,\qd
[\delta_I,\cD_A]=-\Gg IAB\cD_B,\qd
[\delta_I,\delta_J]=\f IJK\delta_K
\label{s2}\eeq
where $[\cdot,\cdot\}$ is the graded commutator,
$\{\cD_A\}$ denotes
collectively the super-covariant derivatives $\cD_a$
and the supersymmetry transformations $\cD_\alpha$ and
$\5\cD_\da$, and $\{\delta_I\}$ contains the
independent elements
of the direct sum $\cG=\cG_L+\cG_{YM}$ of the
Lorentz algebra $\cG_L$ and of the (reductive)
Lie algebra $\cG_{YM}$ of the Yang--Mills gauge group whose
elements are denoted by
$l_{ab}=-l_{ba}$ and $\delta_i$ respectively,
\[
\{\cD_A\}=\{\cD_a,\, \cD_\alpha,\, \5\cD_\da\},\qd
\{\delta_I\}=\{\delta_i,\, l_{ab}:\ a>b\}.
\]
In (\ref{s2}) the $\f IJK$  
are the structure constants
of $\cG$, the $\Gg IAB$ are the entries of the
matrices representing
$\cG$ on the $\cD_A$, and the $\T ABC$ and $\F ABI$ are
torsions and curvatures
spelled out explicitly in appendix \ref{appA}. (\ref{s2})
is nonlinearly realized on tensor fields, i.e.
in old minimal supergravity on 
\beq M,\, \5M,\,  B_a,\, \lambda_\alpha^i,\, \5\lambda_\da^i,\,
\T ab\alpha,\, {T_{ab}}^\da,\, D^i,\, \F abI
\label{s3}\eeq
and all their super-covariant derivatives
($\cD_{a_1}\ldots \cD_{a_k}M$ etc.). The corresponding set 
of tensor fields in new minimal supergravity is obtained by setting
$M$ to zero and using the identification (\ref{identify}).

The BRST transformations of {\em all} the fields can then be
compactly written in the form
\bea
\4s\, \cT&=&(\4\xi^A\cD_A+\4C^I\delta_I)\, \cT,            \label{s5}\\
\4s\, \4\xi^A  &=&\4C^I\Gg IBA \4\xi^B-
                  \s0 12(-)^{\ep_B} \4\xi^B \4\xi^C\T CBA,
                                                        \label{s6}\\
\4s\, \4C^I  &=&
\s0 12\f KJI \4C^J \4C^K-\s0 12(-)^{\ep_A} \4\xi^A \4\xi^B\F BAI,
                                                        \label{s7}\\
\4s\, \4Q&=& \s0 16\,\4\xi^a \4\xi^b \4\xi^c H_{abc}
            + i\,\4\xi^\alpha\4\xi_{\alpha\da}\4\xi^\da
\label{s8}\eea
where $\4s$ is the sum of the BRST operator and the spacetime
exterior derivative $d=dx^\mu\6_\mu$,
\beq \4s=s+d\ ,
\label{tildes}\eeq
$\ep_A$ is the Grassmann parity
of $\cD_A$ ($\ep_a=0$, $\ep_\alpha=\ep_\da=1$), 
$\cT$ denotes an arbitrary tensor field, 
and $\4\xi^A$, $\4C^I$ and $\4Q$ are ``generalized connections"
defined by
\bea
\4\xi^a&=&(C^\mu+dx^\mu)\,\viel a\mu,\nn\\
\4\xi^\alpha&=&\xi^\alpha+(C^\mu+dx^\mu)\,\RS \alpha\mu,\nn\\
\4\xi^\da&=&\5\xi^\da-(C^\mu+dx^\mu)\,\bRS \da\mu,\nn\\
\4C^{ab}&=&C^{ab}+(C^\mu+dx^\mu)\,\spin {ab}\mu,\nn\\
\4C^i&=&C^i+(C^\mu+dx^\mu)\,\A i\mu,\nn\\
\4Q&=&Q+(C^\mu+dx^\mu)\, Q_\mu+
\s0 12(C^\mu+dx^\mu)(C^\nu+dx^\nu)\, t_{\mu\nu}\ .
\label{s9}\eea
It should be noted that the equations (\ref{s5})--(\ref{s8}) 
decompose into parts with different ghost numbers and form degrees.
The reader may check that this decomposition reproduces at 
nonvanishing ghost numbers indeed
the standard BRST transformations for supergravity, such as
(\ref{sviel}) and (\ref{sCdiff})--(\ref{sQ}).
Furthermore the ghost number 0 parts of
(\ref{s5})--(\ref{s8}) provide the explicit form of the
super-covariant derivatives $\cD_a$,
of the spin connection, and of
the super-covariantized field strengths $\T ab\alpha$, $\T ab\da$,
$\F abI$ and $H_{abc}$ respectively (see appendix \ref{appA} 
for details). The consistency of all these formulae and
the nilpotency of the BRST transformations is guaranteed
by the algebra (\ref{s2}) and the Jacobi and Bianchi identities
implied by it.
\medskip

{\em Remark:} I stress again that the $\cD_A$ do {\em not} act in
a superspace here. Rather, they are algebraically (and
nonlinearly) realized on the 
tensor fields which are {\em not} superfields
(see appendix \ref{appA} for more details concerning the approach
used here). One can ``promote" this realization to
superspace but this would not be useful for the cohomological
analysis, see e.g.\ \cite{bp}.

\mysection{Off-shell basis for the tensor fields}\label{basis}

It will be crucial for the analysis to determine
an appropriate ``basis" for the tensor fields, i.e.\ for
the fields (\ref{s3}) and all their super-covariant derivatives.
Here ``basis" is not used in the vector space sense but
for a subset $\{\cT^\TIX\}$ of these tensor fields
which allows to express any tensor field
uniquely in terms of the elements of this subset.
Later, when the full BRST cohomology is computed,
we will construct an analogous on-shell basis. Here we
will construct an off-shell basis%
\footnote{Analogous constructions
can be found in chapters 4 and 5 of \cite{penrose}.}.
To that end we first decompose the torsions
$\T ab\alpha$ and $\T ab\da$ ($\equiv$ gravitino field strengths)
and the curvatures $\F abI$ ($\equiv$ Lorentz- and Yang--Mills
field strengths) into Lorentz irreducible
multiplets, using
$T_{\alpha\da\, \beta\dbe\, \gamma}=
\sigma^a_{\alpha\da}\sigma^b_{\beta\dbe}T_{ab\gamma}$ etc.,
\bea
T_{\alpha\da\, \beta\dbe\, \gamma}&=&
\ep_{\alpha\beta}U_{\da\dbe\gamma}+\ep_{\dbe\da}
(W_{\alpha\beta\gamma}+\s0 23\ep_{\gamma(\alpha}S_{\beta)}),
\label{bas1}\\
F_{\alpha\da\, \beta\dbe}{}^i&=&
\ep_{\alpha\beta}\5G_{\da\dbe}{}^{i}+
\ep_{\da\dbe}G_{\alpha\beta}{}^{i},
\label{bas2}\\
F_{\alpha\da\, \beta\dbe\, \gamma\dg\, \delta{\dot \delta}}&=&
\ep_{\da\dbe}\ep_{\dg{\dot \delta}}[X_{\alpha\beta\gamma\delta}
-\s0 16(\ep_{\alpha\gamma}\ep_{\beta\delta}+
\ep_{\beta\gamma}\ep_{\alpha\delta})\cR]
-\ep_{\alpha\beta}\ep_{\dg{\dot \delta}}Y_{\gamma\delta\da\dbe}
\nonumber\\
& & +\ep_{\alpha\beta}\ep_{\gamma\delta}[\5X_{\da\dbe\dg{\dot \delta}}
-\s0 16(\ep_{\da\dg}\ep_{\dbe{\dot \delta}}+
\ep_{\dbe\dg}\ep_{\da{\dot \delta}})\cR]
-\ep_{\da\dbe}\ep_{\gamma\delta}Y_{\alpha\beta\dg{\dot \delta}}\ .
\label{bas3}\eea
Here the components of $U$, $W$,
$G$, $X$ and $Y$ are
completely symmetric in all their undotted and dotted spinor indices
respectively.
$X$, $Y$ and $\cR$ are the super-covariantized Weyl tensor,
trace-free Ricci tensor
and curvature scalar in spinor notation
respectively (one has $\cR=\F ab{ba}$, 
$Y_{ab}={F_{acb}}^c+\s0 14\eta_{ab}\cR$).

We now construct a basis for all the  super-covariant derivatives
of the tensor fields (\ref{s3}). To that end
we introduce the short hand notation $\cT_n^m$ for a
Lorentz irreducible multiplet of tensor fields whose
components have $m$ dotted and undotted $n$
spinor indices and are completely symmetric in them
respectively,
\[ \cT_n^m\equiv
\{\cT_{\alpha_1\cdots\alpha_n}^{\da_1\cdots\da_m}\},\quad
\cT_{\alpha_1\cdots\alpha_n}^{\da_1\cdots\da_m}
=\cT_{(\alpha_1\cdots\alpha_n)}^{(\da_1\cdots\da_m)}\ .\]
We now define the following operations $\cD_+^+$, \ldots , $\cD_-^-$,
using $\cD_{\alpha\da}=\sigma^a_{\alpha\da}\cD_a$:
\bea \cD_+^+\cT_n^m&\equiv&
\{\cD_{(\alpha_0}^{(\da_0}
\cT_{\alpha_1\cdots\alpha_n)}^{\da_1\cdots\da_m)}\},
\nonumber\\
\cD_+^-\cT_n^m&\equiv&
\{m\, \cD_{\da_m(\alpha_0}
\cT_{\alpha_1\cdots\alpha_n)}^{\da_1\cdots\da_m}\},
\nonumber\\
\cD_-^+\cT_n^m&\equiv&
\{n\, \cD^{\alpha_n(\da_0}
\cT_{\alpha_1\cdots\alpha_n}^{\da_1\cdots\da_m)}\},
\nonumber\\
\cD_-^-\cT_n^m&\equiv&
\{mn\, \cD^{\alpha_n}_{\da_m}
\cT_{\alpha_1\cdots\alpha_n}^{\da_1\cdots\da_m}\} .
\label{bas4}\eea
Using the algebra (\ref{s2})
it is straightforward to verify that these operations satisfy
\bea
[\cD_\pm^\pm,\cD_\pm^\mp]\cT_n^m&=&O(2),\nonumber\\
{}[\cD_\pm^\pm,\cD_\mp^\pm]\cT_n^m&=&O(2),\nonumber\\
{}[\cD_+^+,\cD_-^-]\cT_n^m&=&(m+n+2)\Box \cT_n^m+O(2),
\nonumber\\
{}[\cD_-^+,\cD_+^-]\cT_n^m&=&(m-n)\Box \cT_n^m+O(2),
\nonumber\\
\cD_+^- \cD_-^+\cT_n^m&=&\s0 12\,n(m+2)\, \Box\cT_n^m
+\cD_+^+\cD_-^-\cT_n^m+O(2).
\label{bas5}\eea
Here $O(2)$ denotes terms which are at least quadratic in the
tensor fields and
we used the super-covariant d'Alembertian
\[ \Box=\cD_a\cD^a=\s0 12 \cD_{\alpha\da}\cD^{\da\alpha}\ . \]
The relations (\ref{bas5}) imply that, up to
terms $O(2)$, all super-covariant derivatives of an arbitrary
$\cT_n^m$ can be expressed as linear combinations of
the following quantities:
\beq \Box^p(\cD_+^+)^q(\cD_+^-)^r(\cD_-^-)^s\cT_n^m\ ,\quad
\Box^p(\cD_+^+)^q(\cD_-^+)^{r'}(\cD_-^-)^{s'}\cT_n^m
\label{bas6}\eeq
where $p,q\geq 0$, $(r+s)\leq m$, $s\leq n$,
$(r'+s')\leq n$, $s'\leq m$
and $r'>0$ (the last condition avoids a double counting
of the terms with $r=0$ and $r'=0$).
Note that none of the expressions  (\ref{bas6}) contains both
$\cD_+^-$ and $\cD_-^+$ as a consequence of the last two relations in
(\ref{bas5}).

Up to terms $O(2)$, (\ref{bas6}) provides already a basis for the
super-covariant derivatives of the `elementary' (non-composite) 
tensor fields $M$, $B$, $\lambda^i$ and $D^i$.
However, when applied to the field strengths, it
yields an overcomplete set of super-covariant derivatives because there
are still algebraic identities relating the elements of this set.
These remaining identities reflect the Bianchi identities
\beq
\cD_{[a}\T b{c]}\alpha=O(2),\quad
\cD_{[a}\F b{c]}I=O(2)
\label{bas7}\eeq
and the super-covariant derivatives thereof.
In order to remove this remaining redundancy, we
first decompose the identities (\ref{bas7}) into
Lorentz irreducible parts. In the above notation the result is
\bea
& &\cD_-^- U+2\cD_-^+S=O(2),\label{bas8}\\
& &2\cD_-^+ W+2\cD_+^+S-3\cD_+^-U=O(2),\label{bas9}\\
& &\cD_-^+X-2\cD_+^-Y=O(2),\label{bas10}\\
& &\cD_-^-Y-2\cD_+^+\cR=O(2),\label{bas11}\\
& &\cD_-^+G^i+\cD_+^-\5G^i=O(2)\label{bas12}
\eea
and the complex conjugation of (\ref{bas8})--(\ref{bas10})
(the remaining identities (\ref{bas11}) and (\ref{bas12}) are real).
Now, (\ref{bas8}), all its super-covariant derivatives and the 
corresponding complex conjugated identities show that
all super-covariant derivatives (\ref{bas6}) of $U$ and $\5U$ containing
the  operation $\cD_-^-$ can be expressed in terms of
super-covariant derivatives
of $S$ and $\5S$ up to terms $O(2)$. Analogously
(\ref{bas9})--(\ref{bas12}) (and their super-covariant derivatives
and complex conjugations) allow to
eliminate all super-covariant derivatives (\ref{bas6})
of $W$, $\5W$, $X$, $\5X$ and $\5G^i$ except for those
with $(p,q,r,s)=(0,q,0,0)$,
and all super-covariant derivatives
of $Y$ containing the operation $\cD_-^-$, up to terms of
$O(2)$. As the $O(2)$ terms are composed
of tensor fields that have lower dimension than the 
respective considered
quantity, one can easily prove by induction that
in old minimal supergravity the following list
provides  a basis $\{\cT^\TIX\}$ for the tensor fields
in the above sense:
\bea
M,\5M:& &
\Box^p(\cD_+^+)^q\{ M,\, \5M\};\nonumber\\
B: & &\Box^p(\cD_+^+)^q\{B,\, \cD_+^-B,\, \cD_-^+B,\, \cD_-^-B\};
\nonumber\\
S,\5S:& &\Box^p(\cD_+^+)^q
\{S,\, \5S,\, \cD_-^+S,\, \cD_+^-\5S\};\nonumber\\
U,\5U:& &\Box^p(\cD_+^+)^q
\{U,\, \5U,\, \cD_-^+U,\, \cD_+^-U,\,
\cD_-^+\5U,\, \cD_+^-\5U,\,
(\cD_+^-)^2U,\, (\cD_-^+)^2\5U\};\nonumber\\
W,\5W:& &
(\cD_+^+)^q\{ W,\, \5W\};\nonumber\\
\cR:& &\Box^p(\cD_+^+)^q\cR;\nonumber\\
Y:& &\Box^p(\cD_+^+)^q\{Y,\, \cD_-^+Y,\, \cD_+^-Y,\,
(\cD_-^+)^2Y,\, (\cD_+^-)^2Y\};\nonumber\\
X,\5X:& &(\cD_+^+)^q\{ X,\, \5X\};\nonumber\\
\lambda^i,\5\lambda^i:& &
\Box^p(\cD_+^+)^q
\{\lambda^i,\, \5\lambda^i,\,
\cD_-^+\lambda^i,\, \cD_+^-\5\lambda^i\};\nonumber\\
G^i,\5G^i:& & (\cD_+^+)^q\5G^i,\quad
\Box^p(\cD_+^+)^q\{G^i,\, \cD_-^+G^i,\, (\cD_-^+)^2G^i\}
;\nonumber\\
D^i:& &\Box^p(\cD_+^+)^qD^i\label{bas13}\eea
where $p,q=0,1,\ldots$ and we used the notation
\[\Box^p(\cD_+^+)^q\{ M,\, \5M\}\equiv
\{\Box^p(\cD_+^+)^q M,\, \Box^p(\cD_+^+)^q \5M\}\quad etc.\ .\]
I stress that all the tensor fields listed in (\ref{bas13}) 
are algebraically independent, i.e. a function
$f(\cT)$ vanishes off-shell if and only if it vanishes
identically in terms of the elementary fields and their
derivatives.

An analogous off-shell basis for the tensor fields 
in new minimal supergravity
is obtained from the above list by discarding $M$, $\5M$ and $B$
(and, if present, the tensor fields associated with $R$-transformations)
and adding to it
\bea  
& &\Box^p(\cD_+^+)^q\{\4H,\, \cD_+^-\4H,\, \cD_-^+\4H\},
\nonumber\\
& &(\cD_+^+)^q\5G^\dR,\quad 
\Box^p(\cD_+^+)^q\{G^\dR,\, \cD_-^+G^\dR,\, (\cD_-^+)^2G^\dR\}
\label{bas14}\eea
where $\4H\equiv\{\4H_\alpha^\da\}$ denotes
the dual of $H_{abc}$, $\4H^a=\s0 16\ep^{abcd}H_{bcd}$
(one has $\cD_-^-\4H=0$ and therefore (\ref{bas14})
contains no terms with $\cD_-^-$).

\mysection{Restricted BRST cohomology}
\label{rest}

\subsection{Reduction to the covariant cohomology of $\4s$}
\label{reduce}

To compute the restricted BRST cohomology we have to solve
(\ref{i1}) and (\ref{i1b}) for antifield independent local forms
$\omega_4$, $\omega_3$, $\eta_4$ and $\eta_3$.
The first step towards the solution of this problem
reduces it (locally) to the cohomology of $\4s=s+d$
on local functions
depending only on the tensor fields $\cT^\TIX$
given in the previous section
and on the generalized connections (\ref{s9}).
To show this we use the fact that (\ref{i1}) implies
descent equations for $s$ and $d$,
\beq s\omega_p+d\omega_{p-1}=0,\quad p=0,\ldots,4\quad
(\omega_{-1}\equiv 0),
\label{deseqs}\eeq
which are compactly written in the form
\beq \4s\, \omega=0,\quad \omega=\sum_{p=0}^4\omega_p\ .\label{i2}\eeq
The nontriviality condition (\ref{i1b})
translates into
\beq \omega\neq\4s\,\eta+constant\label{i2a}\eeq
where $\eta$ is a sum of local forms like $\omega$.
Such sums are the {\em local total forms} mentioned in the 
introduction.
Thanks to the closure of the gauge algebra, $\4s$ is strictly
nilpotent on antifield independent forms and therefore the
restricted local BRST cohomology can be obtained
from the cohomology of $\4s$ in the space of antifield
independent local total forms.

The analysis can be considerably simplified by
a suitable choice of local ``jet coordinates"
in terms of which all antifield independent total forms
can be expressed, cf.\ \cite{ten} for details. An appropriate
set of local jet coordinates is $\{\cU^l,\cV^l,\cW^i\}$ where
\bea
\{\cU^l\}&=&\{x^\mu,\,
\6_{(\mu_1\ldots\mu_{k}}\viel a{\mu_{k+1})},\,
\6_{(\mu_1\ldots\mu_k}\omega_{\mu_{k+1})}{}^{ab},  \nonumber\\
& &\phantom{\{}
\6_{(\mu_1\ldots\mu_k}\psi_{\mu_{k+1})}{}^\alpha,\,
\6_{(\mu_1\ldots\mu_k}\5\psi_{\mu_{k+1})}{}^{\dot \alpha},\nonumber\\
& &\phantom{\{}
\6_{(\mu_1\ldots\mu_k}t_{\mu_{k+1})\nu},\,
\6_{(\mu_1\ldots\mu_k}Q_{\mu_{k+1})},\nonumber\\
& &\phantom{\{}
\6_{(\mu_1\ldots\mu_{k}}\A i{\mu_{k+1})}:\ k=0,1,\ldots \},\label{Us}\\
\{\cV^l\}&=&\{\4s\, \cU^l\}\label{Vs}\\
\{\cW^i\}&=&\{\4C^I,\, \4\xi^A,\, \4Q,\, \cT^\TIX\}.\label{Ws}
\eea
These jet coordinates satisfy
\beq \4s\, \cU^l=\cV^l,\quad \4s\, \cW^i=R^i(\cW)\label{r0}\eeq
which implies by K\"unneth's formula that the
$\4s$-cohomology  on local total forms $\omega(\cU,\cV,\cW)$
can be obtained from the $\4s$-cohomologies
on local total forms  $\alpha(\cU,\cV)$ and $\alpha(\cW)$
according to
\beq \4s\, \omega(\cU,\cV,\cW)=0\ \then\
\omega(\cU,\cV,\cW)=a_{ai}\,\alpha^i(\cU,\cV)\, \alpha^a(\cW)
+\4s\,\eta( \cU,\cV,\cW).\label{kunneth}\eeq
Here the $a_{ai}$ are constants
and $\alpha^i(\cU,\cV)$ and $\alpha^a(\cW)$ represent
nontrivial $\4s$-cohomology classes in the space
of local total forms $\alpha(\cU,\cV)$ and $\alpha(\cW)$ respectively.

Now, in the manifold of the $\cU$'s and $\cV$'s the $\4s$-cohomology
is {\em locally} trivial thanks to (\ref{r0}) and thus
locally represented
by a constant. Hence, locally the restricted BRST cohomology
reduces to the $\4s$-cohomology on local total
forms $\alpha(\cW)$ and can be obtained from the solutions of
\beq \4s \alpha(\cW)=0,\qd
\alpha(\cW)\neq
\4s \beta(\cW)+constant.\label{r2}\eeq
The modifications due to global (topological) aspects of the
manifold of the $\cU$'s and $\cV$'s
are roughly sketched in section \ref{top} along
the lines of \cite{bbhgrav}.

For later purpose I add a few comments.
\ben
\item If $\omega_4$ has ghost number $g$,
then $\omega$ has {\em total degree} $G=g+4$ where
the total degree ($\tot$) is given by the sum of the
form degree ($\deg$) and the ghost number ($\gh$),
\beq \tot=\gh+\deg .\label{tot}\eeq
Hence, the (restricted) BRST cohomology at ghost number $g$
is obtained from the (restricted) $\4s$-cohomology
at total degree $G=g+4$.
The total degrees of the $\cW$'s are given by
\beq \tot(\cT^\TIX)=0,\qd \tot(\4\xi^A)=\tot(\4C^I)=1,\qd
\tot(\4Q)=2.\label{r55}\eeq

\item We can assume without loss of generality that
the solutions to (\ref{r2}) are real. Indeed, as $\4s$ is
a real operator (see appendix \ref{app2} for the conventions
concerning complex conjugation used here), $\4s\alpha=0$
implies that both the real and the imaginary
part of $\alpha$ are separately $\4s$-invariant, and
$\alpha=\4s\beta$ implies that both parts are $\4s$-exact.

\item (\ref{r0}) implies that $\alpha(\cW)$ is $\4s$-exact
in the space of local total forms $\alpha(\cW)$ whenever
it is $\4s$-exact in the space of all local total forms,
\beq \alpha(\cW)=\4s\, \eta(\cU,\cV,\cW)\quad\then\quad
\exists\, \beta:\ \alpha(\cW)=\4s\, \beta(\cW).\label{r33}\eeq

\item In the following  the analysis will be carried
out for old and new minimal supergravity simultaneously.
It should be kept in mind however that $\4Q$ and the super-covariant
field strength $H_{abc}$ of $t_{\mu\nu}$
occur only in new minimal supergravity,
whereas the auxiliary fields $M$ and $B_a$ are present only in old
minimal supergravity.

\item The space of local total forms $\alpha(\cW)$ to be considered
for the computation of the restricted $\4s$-cohomology 
is actually the space of polynomials in the $\cW$'s.
This follows from the fact that $\4s$ has
total degree 1 and vanishing
dimension on the $\cW$'s. Hence, the restricted
$\4s$-cohomology 
can be computed separately for each subspace of
total forms with given dimension and total degree.
As all tensor fields have positive dimension,
{\em local} total forms $\alpha(\cW)$ with fixed
dimension and total degree
are necessarily polynomials
in the $\cW$'s since we do not allow them to involve negative
powers of the auxiliary fields (because otherwise
they may become nonlocal or even ill-defined after
eliminating the
auxiliary fields through their equations of motion)%
\footnote{All the arguments go through also for
formal local series' in the $\cW$'s. This can become important
when matter multiplets are included because then
it might not be justified anymore to restrict the investigation
to polynomials in the $\cW$'s.}.
\een

\subsection{Decomposition of the problem}\label{decomp}

To solve (\ref{r2}), we expand it in $\4Q$ and in the $\4C^I$,
using
\beq \alpha(\cW)=\sum_{k=0}^m
\alpha_k(\cW),\qd
N_{CQ}\,\alpha_k=k\,\alpha_k
\label{r3}\eeq
where 
$N_{CQ}$ is the counting operator for the $\4C^I$ and $\4Q$,
\beq
N_{CQ}=\4C^I\,\frac{\6}{\6\4C^I}+\4Q\,\frac{\6}{\6\4Q}\ .
\label{r4}\eeq
Note that $m$ in (\ref{r3}) cannot exceed the total degree of $\alpha$
because $\alpha$ does not involve antifields
(hence, $m$ is always finite).
$\4s$ decomposes on the $\cW$'s
into three parts,
\beq \4s\, \alpha(\cW)=
(\4s_{lie}+\4s_{susy}+\4s_{curv})\,\alpha(\cW),
\label{r5}\eeq
which have $N_{CQ}$-degrees $1,0,-1$ respectively,
\beq [N_{CQ},\4s_{lie}]=\4s_{lie}\, ,\qd
[N_{CQ},\4s_{susy}]=0\, ,\qd
[N_{CQ},\4s_{curv}]=-\4s_{curv}\ .
\label{r6}\eeq
Note that this implies the following anticommutation relations
due to $\4s^2=0$:
\bea
& &(\4s_{lie})^2=(\4s_{curv})^2=\{\4s_{lie},\4s_{susy}\}=
\{\4s_{curv},\4s_{susy}\}=0,\nonumber\\
& &\{\4s_{lie},\4s_{curv}\}+(\4s_{susy})^2=0.
\label{r7}\eea
The three parts of $\4s$ are spelled out explicitly in table 4.1
where $\cF^I$ and $\cH$ are total super-curvature forms given by
\bea \cF^I&=&
-\s0 12(-)^{\ep_A} \4\xi^A \4\xi^B\F BAI,
\label{r6a}\\
\cH&=&\s0 16\,\4\xi^a \4\xi^b \4\xi^c H_{abc}
      + i\,\4\xi^\alpha\4\xi_{\alpha\da}\4\xi^\da\ .
\label{r6b}\eea
It is evident from table 4.1 that the three parts of
$\4s$ have well-distinguished interpretations
indicated by their subscripts:
$\4s_{lie}$ encodes the Lie algebra of $\cG$,
$\4s_{susy}$ contains the supersymmetry transformations
and the super-covariant derivatives,
and $\4s_{curv}$ transforms the $\4C^I$ and $\4Q$ into the
corresponding total super-curvature forms.
\[
\ba{c|c|c|c}
\cW            &  \4s_{lie}\cW & \4s_{susy}\cW & \4s_{curv}\cW\\
\hline\rule{0em}{3ex}
\4Q          &   0 & 0  &     \cH  \\
\rule{0em}{3ex}
\4C^I        &  \s0 12\4C^J\4C^K\f KJI & 0 & \cF^I\\
\rule{0em}{3ex}
\4\xi^A      &  \4C^I\Gg IBA\4\xi^B
             & -\s0 12 (-)^{\ep_B}\4\xi^B\4\xi^C\T CBA & 0 \\
\rule{0em}{3ex}
\cT     & \4C^I \delta_I\cT & \4\xi^A\cD_A\cT & 0 \\
\multicolumn{4}{c}{}\\
\multicolumn{4}{c}{\mbox{Table 4.1: Decomposition of $\4s$}}
\ea
\]
$\4s \alpha=0$ thus decomposes into
\bea 0&=& \4s_{lie} \alpha_m\ ,\label{r8a}\\
0&=&\4s_{susy}\alpha_m+\4s_{lie}\alpha_{m-1}\ ,\label{r8b}\\
0&=&\4s_{curv}\alpha_m+\4s_{susy}\alpha_{m-1}+\4s_{lie}\alpha_{m-2}\ ,
\label{r8c}\\
&\vdots&\nonumber\eea
Now, the cohomology of $\4s_{lie}$ is well-known. It is just
the Lie algebra cohomology of $\cG$.  Since
all the $\4\xi^A$ and the tensor fields transform according to finite
dimensional linear representations of $\cG$ and since
$\cG$ is reductive by assumption, the
cohomology of $\4s_{lie}$ is represented by functions of the
form $ f_i(\4\xi,\cT)P^i(\Th,\4Q)$ where the $f_i$ are
$\cG$-invariant and
the $P^i(\Th,\4Q)$ are linearly independent monomials in $\4Q$ and
in the primitive elements $\Th_K$ of the
Lie algrebra cohomology of $\cG$.
The $\Th_K$ themselves are polynomials in the $\4C^I$ and
correspond to the independent Casimir operators of $\cG$.
Their number therefore equals the rank of $\cG$
($K=1,\ldots,rank(\cG)$). They can be constructed
using appropriate matrix representations $\{T^{(K)}_I\}$ of $\cG$
(see e.g. \cite{lie}),
\beq
\Th_K=(-)^{m_K+1}\,\frac {m_K!(m_K-1)!}{(2m_K-1)!}
         \, Tr(\4C^{2m_K-1}),\quad \4C=\4C^I T^{(K)}_I
\label{thetas}\eeq
where $m_K$ denotes the order of the Casimir operator
corresponding to $\Th_K$.

(\ref{r8a}) implies thus
\beq \alpha_m= f_i(\4\xi,\cT)\, P^i(\Th,\4Q),
\quad \delta_I f_i(\4\xi,\cT)=0
\label{r9}\eeq
where
\beq \delta_I \4\xi^A=\Gg IBA\4\xi^B.
\label{r10a}\eeq
In (\ref{r9}) a contribution to $\alpha_m$
of the form $\4s_{lie} \beta_{m-1}$ has been neglected 
without loss of generality because such a contribution
can be always removed from $\alpha_m$ by subtracting the trivial
piece $\4s \beta_{m-1}$ from $\alpha$.

Inserting (\ref{r9}) in (\ref{r8b}) results in
$ (\4s_{susy}f_i)P^i+\4s_{lie}\alpha_{m-1}=0$. Using the
Lie algebra cohomology again, we conclude
\beq \4s_{susy}f_i(\4\xi,\cT)=0.\label{r11}\eeq
Furthermore, we can assume without loss of generality
that none of the $f_i$ is of the form $\4s h(\4\xi,\cT)$.
Namely otherwise we could remove that particular $f_i$
by subtracting the trivial piece $\4s (hP^i)$ from $\alpha$.
Notice that such a subtraction does not clash with the
other redefinitions made so far. In particular it does not
reintroduce a term $\4s_{lie} \beta_{m-1}$ in (\ref{r9})
(note that in general a similar reasoning would {\em not} apply 
if $h$ would depend on one of the $\4C^I$).

Hence, the $f_i$ are determined by the
cohomology of $\4s_{susy}$ in the space of $\cG$-invariant 
local total forms depending only on the $\4\xi$ and $\cT$
(note that $\4s_{susy}$ is nilpotent in this space).  
Now, the latter cohomology is completely
equivalent to the $\4s$-cohomology in the space of local total
forms $f(\4\xi,\cT)$.
Indeed (\ref{r9}) and (\ref{r11}) require $f_i$
to be $\4s$-invariant because $\delta_If_i=0$ implies $\4s_{lie}f_i=0$
and because $\4s_{curv}f_i=0$ holds trivially
as $f_i$ depends neither on the $\4C^I$ nor on $\4Q$.
Conversely, $\4s f_i(\4\xi,\cT)=0$ 
requires $f_i$ to be $\cG$-invariant 
and $f_i(\4\xi,\cT)=\4s h(\4\xi,\cT)$ requires $h$ to be 
$\cG$-invariant (this is evident from expanding these equations
in $\4C^I$). The $f_i$ can thus be assumed to solve
\beq \4s f_i(\4\xi,\cT)=0,\qd f_i(\4\xi,\cT)\neq
\4s h_i(\4\xi,\cT).\label{r12}\eeq
Note that this does not necessarily imply that $f_i$ solves
(\ref{r2}) too. Namely one might have $f_i=\4s\beta(\cW)$
but nevertheless $f_i\neq \4s h(\4\xi,\cT)$.

\subsection{Cohomology of $\4s$ on local total forms $f(\4\xi,\cT)$}
\label{susy}

The solution of
(\ref{r12}) is the most involved part and the
cornerstone of the computation
of the restricted BRST cohomology. 
The result is the following:
\msk

Nonconstant solutions
to (\ref{r12}) exist only at total degrees 2, 3 and 4,
\bea
& &\4sf(\4\xi,\cT)=0,\qd \tot(f)=G\nn\\
& &\LRA\qd f(\4\xi,\cT)=
\left\{\ba{ll}
constant & \mbox{for $G=0$}\\[.5ex]
\4sh(\cT) & \mbox{for $G=1$}\\[.5ex]
a_{i_a}\cF^{\,i_a}+\4sh(\4\xi,\cT) & \mbox{for $G=2$}\\[.5ex]
a\, \cH +\4sh(\4\xi,\cT)& \mbox{for $G=3$}\\[.5ex]
a^\Delta \cP_\Delta+\4sh(\4\xi,\cT)& \mbox{for $G=4$}\\[.5ex]
\4sh(\4\xi,\cT)& \mbox{for $G>4$.}
\ea\right.
\label{c3}\eea
Here the $a$'s are constants, the $h$'s are $\cG$-invariant,
the $\cF^{\,i_a}$ are the
{\em abelian} total curvature forms
(\ref{r6a}), $\cH$ is the total curvature form
corresponding to $t_{\mu\nu}$
given in (\ref{r6b}), and the $\cP_\Delta$ are
reminiscent of superspace integrals,
\beq
\cP_\Delta =(\7\cD_\da \7\cD^\da-M)\, \Xi\,
\{ \cA_\Delta(\5M,\5W,\5\lambda)+
(\cD^\alpha \cD_\alpha-\5M) \cB_\Delta(\cT)\}+c.c.
\label{cp13b}\eeq
where
\bea & &\Xi=-\s0 1{24}\, \varepsilon_{abcd}\,
\4\xi^a\4\xi^b\4\xi^c\4\xi^d\ ,\label{cp14}\\
& &\7\cD_\da = \4\xi^B\T B\da{A}
\frac{\6}{\6\4\xi^A}+\5\cD_\da\ ,
\label{cp15b}\\
& &\delta_I\cP_\Delta=0.\label{cp15c}\eea
In (\ref{cp13b}) the operators $\7\cD_\da$ act on everything
on their right (i.e.\ $\7\cD_\da \7\cD^\da$
acts on $\Xi \{\cdots\}$, not just
on the ``total volume form" $\Xi$).
The $\T ABC$ in (\ref{cp15b})  are the torsions of the
supergravity algebra (\ref{s2}) spelled out explicitly in
appendix \ref{appA}. 
As indicated by its arguments,
$\cA_\Delta$ is a polynomial in the
{\em undifferentiated} fields
$\5M,\, \5W_{\da\dbe\dg},\, \5\lambda^i_\da$ 
($W_{\alpha\beta\gamma}$ is the chiral part of
the gravitino field strength, see section \ref{basis}).
(\ref{cp15c}) requires
$\cB_\Delta$ to be $\cG$-invariant and
$\cA_\Delta$ to be $\cG$-invariant 
except under $R$-transformations
(if $R$-transformations
are gauged, $\cA_\Delta$  must have $R$-charge
$-2$ as $\7\cD_\da$ carries
$R$-weight 1 according to the conventions used here,
see appendix \ref{appA}). I stress that the
$\cA_\Delta$ and $\cB_\Delta$ need not satisfy
any additional requirement in order to ensure
$\4s\cP_\Delta=0$ (but of course we require
them to be local functions).
The term $a\cH$ occurring in (\ref{c3}) for $G=3$
contributes only in new minimal supergravity and
in (\ref{cp13b}) the field $M$ has to be set to zero
in new minimal supergravity.

In the remainder of this section I sketch the
basic ideas used to derive the above result, relegating
details of the computation to the appendices. Let us
first consider the cases $G<4$. In these cases we can
use the fact that {\em any} $\4s$-closed
local total form that does not depend on antifields (!) is
necessarily (locally) $\4s$-exact in the space of
antifield independent local total forms except for the
constants,
\bea & &\4s\, \omega(\cU,\cV,\cW)=0,\
\tot(\omega)=G<4
\nonumber\\
& &\then\quad \omega=\left\{
\ba{ll}
constant         & \mbox{for}\ G=0\\
\4s \eta(\cU,\cV,\cW) & \mbox{for}\ 0<G<4.
\ea\right.
\label{c2}\eea
This follows from very general arguments
which are not restricted to the theories under study but hold 
analogously whenever the gauge algebra closes off-shell.
Indeed, recall that $\4s \omega=0$ decomposes
into descent equations (\ref{deseqs}) and that the
nontrivial (and nonconstant) solutions of these equations
in the space of antifield independent local forms correspond
(locally) one-to-one to the nontrivial solutions  of (\ref{i1})
in the space of antifield independent local volume forms.
Now, if $\omega$ has total degree $<4$ and does not 
depend on antifields, it does not
contain a nontrivial solution to (\ref{i1}), simply
because it cannot contain a
volume form. This implies (\ref{c2}).

In particular we conclude from (\ref{c2}) that
$\4s f(\4\xi,\cT)=0$ implies $f=\4s \eta$ for $0<G<4$ and
$f=constant$ for $G=0$.
However, this does not yet solve our
problem (\ref{r12}) for $G<4$
because to that end we have still to investigate whether or not
$\eta$ can be assumed to depend only on the $\4\xi$ and $\cT$.
This investigation is performed in
appendix \ref{CPL} within a derivation of the
``super-covariant Poincar\'e lemma" which is the
counterpart of the
``covariant Poincar\'e lemma'' proved in \cite{grav}
for standard gravity coupled to Yang--Mills fields\footnote{See also
\cite{Gilkey,com,violette} for analogous results in standard
gravity and Yang--Mills theory.}.
The result is  that  one can indeed choose
$\eta$ to be of the form $h(\4\xi,\cT)$
except that in the case $G=2$ it
may involve a linear combination of the abelian $\4C$'s and
in the case $G=3$ it may contain a term proportional to $\4Q$.
This provides the results for $G<4$ in (\ref{c3})
due to $\4s\4C=\cF$ for the abelian $\4C$'s and
$\4s\4Q=\cH$.

The investigation of (\ref{r12}) is much more involved in the
cases $G\geq 4$. In particular the supersymmetric structure
of the theory plays an essential role in these cases.
The strategy to attack this problem is based on spectral
sequence techniques using the degree in the tensor fields
as filtration. That is to say, we decompose the equation
$\4s f(\4\xi,\cT)=0$ into
parts with definite degree in the tensor fields and analyze it
starting from the part with lowest degree.
The decomposition is unique and thus well-defined thanks to
the completeness and algebraic independence of the $\cT$'s,
cf.\ section \ref{basis}.
The decomposition of $\4s$ takes the form
\beq \4s=\sum_{k\geq 0}\4s_{\tdeg k}\ ,\qd
[N_{\cT},\4s_{\tdeg k}]=k\4s_{\tdeg k}\label{c4}\eeq
where $N_{\cT}$ is the counting operator for the $\cT$'s,
\beq N_{\cT}=\cT^\TIX\frac{\6}{\6\cT^\TIX}\ .\label{c5a}\eeq
$f(\4\xi,\cT)$ decomposes into
\beq f(\4\xi,\cT)=\sum_{k\geq \ell}f_{\tdeg k}(\4\xi,\cT),\qd
N_{\cT}f_{\tdeg k}(\4\xi,\cT)=kf_{\tdeg k}(\4\xi,\cT).
\label{c5}\eeq
$\4s f=0$ requires, at lowest order in the tensor fields,
\beq \4s_{\tdeg 0} f_{\tdeg \ell}(\4\xi,\cT)=0.
\label{c6}\eeq
Furthermore we can
remove any piece of the form
$\4s_{\tdeg 0} h_{\tdeg \ell}(\4\xi,\cT)$ from $f_{\tdeg \ell}$
without changing the cohomology class of $f$
by substracting $\4s h_{\tdeg \ell}$ from $f$.
Hence, $f_{\tdeg \ell}$ is
actually determined by the $\4s_{\tdeg 0}$-cohomology in
the space of local total forms $f(\4\xi,\cT)$.
In particular we can assume
\beq f_{\tdeg \ell}\neq \4s_{\tdeg 0} h_{\tdeg \ell}(\4\xi,\cT).
\label{c7}\eeq

The analysis of (\ref{c6}) and
(\ref{c7}) proceeds along the lines of an analogous
investigation performed in \cite{glusy} for rigid supersymmetry.
This is possible thanks to the structure of $\4s_{\tdeg 0}$
which splits into
\beq \4s_{\tdeg 0}=\4s_{lie}+\delta_{susy}\ .\label{c9}\eeq
Here $\delta_{susy}$ is
nothing but the linearized part of $\4s_{susy}$
(see table 4.1) and acts according to
\bea \delta_{susy}\4\xi^a&=&
2i\, \4\xi^\alpha\4\xi^\da\sigma^a_{\alpha\da}\ ,
\label{c8}\\
\delta_{susy}\4\xi^\alpha&=&
\delta_{susy}\4\xi^\da\ =\ 0,\label{c9a}\\
\delta_{susy}\cT^\TIX&=&\4\xi^A D_A \cT^\TIX
\label{c9b}\eea
where $D_A\cT$ is that part of $\cD_A\cT$ which is
linear in the $\cT$'s. The $D_A$ are thus {\em linearly}
realized on the tensor fields, in contrast to the $\cD_A$,
and satisfy an algebra involving only structure constants
(rather than field dependent structure functions). This algebra is
the linearized version of the covariant supergravity algebra
of the $\cD_A$. It reads simply
\beq
\{D_\alpha,\5D_\da\}=-2i\sigma^a_{\alpha\da}D_a\ ,
\quad [D_A,D_B\}=0\quad \mbox{otherwise}.
\label{c10}\eeq
(\ref{c6}) requires $f_{\tdeg \ell}$ in particular to be $\cG$-invariant
in consequence of the presence of $\4s_{lie}$ in $\4s_{\tdeg 0}$
(recall that the $\4s$-cohomology on total forms $f(\4\xi,\cT)$
is nothing but the $\4s_{susy}$-cohomology on 
$\cG$-invariant total forms $f(\4\xi,\cT)$, 
see section \ref{decomp}).
Hence, (\ref{c6}) and (\ref{c7}) are equivalent to
\beq \delta_{susy}f_{\tdeg \ell}(\4\xi,\cT)=0,\quad
f_{\tdeg \ell}\neq \delta_{susy}h_{\tdeg \ell}(\4\xi,\cT),
\quad \delta_If_{\tdeg \ell}=\delta_Ih_{\tdeg \ell}=0.
\label{c99}\eeq
$f_{\tdeg \ell}$ is thus determined by the
$\delta_{susy}$-cohomology on $\cG$-invariant
local total forms $f(\4\xi,\cT)$. 
This problem is indeed analogous to the one investigated
in \cite{glusy} because
(\ref{c10}) is of course nothing but the familiar algebra
of rigid supersymmetry analyzed there.
The only difference to the investigation performed in
\cite{glusy} is that in our case (\ref{c10}) is represented on tensor 
fields whereas in \cite{glusy} it was represented on ordinary
fields where $D_a$ reduces to $\6_a$. This difference does
not prevent us from using the methods of \cite{glusy}. Namely
the only property 
needed to adopt the analysis and results of \cite{glusy}
is that the representation of the
subalgebra $\{D_\alpha,D_\beta\}=0$ of (\ref{c10}) has
{\em QDS-structure} in the terminology of \cite{glusy},
both in old and in new minimal supergravity.
This is explained in detail in appendix \ref{appqds}. 
It is quite remarkable that this property alone
allows us to solve first (\ref{c99})
and then (\ref{r12}) completely in the cases $G\geq 4$.

Indeed, as in section 6 of \cite{glusy}%
\footnote{See also appendix \ref{global} of the present paper
where the derivation of the
analogous result for the on-shell cohomology is sketched.} one
proves by means of the QDS-structure of minimal
supergravity that all the solutions to (\ref{c6}) with
$G> 4$ are $\4s_{\tdeg 0}$-exact, while
the nontrivial solutions with $G=4$ are linearized
counterparts of (\ref{cp13b}) given by
\beq
P_\Delta =\7D_\da \7D^\da \, \Xi\,
\{ \cA_\Delta(\5M,\5W,\5\lambda)+
D^\alpha D_\alpha \cB_\Delta(\cT)\}+c.c.
\label{restr1}\eeq
with $\Xi$ as in (\ref{cp14}) and
\beq
\7D_\da = 2i\4\xi^\alpha\sigma^a_{\alpha\da}
\frac{\6}{\6\4\xi^a}+\5D_\da\ .\label{cp15a}
\eeq
The results (\ref{c3}) for $G\geq 4$ follow then
from standard arguments of spectral
sequence techniques and from the fact that
any $\cG$-invariant local total form (\ref{cp13b}) is indeed
$\4s$-invariant as shown in \cite{sugra}.

\subsection{Completion of the analysis}\label{completion}

The results of sections \ref{decomp} and
\ref{susy} imply that any solution
to $\4s\alpha(\cW)=0$ is, up to a trivial solution $\4s\beta(\cW)$,
an $\4s$-invariant completion of
a local total form
\beq f_i(\4\xi,\cT)\, P^i(\Th,\4Q),\quad
f_i(\4\xi,\cT)\in\{1,\,\cF^{\, i_a},\,
\cH,\, \cP_\Delta\}.
\label{comp1}\eeq
We are therefore left with the following problem:
which total forms (\ref{comp1}) have a
local $\4s$-invariant completion and which of these completions are
inequivalent in the restricted BRST cohomology?

To answer these questions I show first that each
of the following local total forms can be completed
to a local solution of $\4s\alpha=0$:
\beq
P(\Th)+\cH \7P(\Th,\4Q)+ \cP_\Delta P^\Delta(\Th,\4Q)
\label{comp3}\eeq
where $P^\Delta$ and $\7P$ are arbitrary polynomials in
$\4Q$ and the $\Th_K$, whereas $P$ depends only
on those $\Th_K$ with $m_K>2$ (cf. (\ref{thetas})),
\beq \frac{\6P(\Th)}{\6\Th_K}=0
\quad \mbox{for}\quad m_K=1,2.
\label{comp3a}\eeq
Note that the second term in (\ref{comp3}) contributes only in
new minimal supergravity.

In fact each term in (\ref{comp3}) can separately
be completed to a solution to $\4s\alpha=0$.
To show this it is useful to complete first $\Th_K$
to a ``total super-Chern--Simons form" $q_K$, analogously
to the standard construction in Yang--Mills
theory (see e.g.\ \cite{StoraZumino}),
\bea & &q_K=m_K\int_0^1 dt\, Tr\left[\4C
\cF_t^{m_K-1}\right],\nonumber\\
& & \cF_t=t\cF+(t^2-t)\4C^2,\quad
\4C=\4C^IT^{(K)}_I,\quad
\cF=\cF^IT^{(K)}_I\label{comp4}\eea
with $T^{(K)}_I$
as in (\ref{thetas}). Due to (\ref{s7}), resp.
\[ \4s\4C+\4C^2=\cF, \]
the $q_K$ satisfy
\beq \4s\, q_K=Tr(\cF^{m_K})\equiv f_K.\label{comp5}\eeq
The $f_K$ are of course the supersymmetric counterparts of
the familiar characteristic classes. Note however that
they do {\em not} necessarily vanish for $m_K>2$ because they
are total forms
decomposing into local differential forms with various form degrees.
Hence, the $q_K$ are not automatically $\4s$-invariant for $m_K>2$,
in contrast to their counterparts in Yang--Mills theory and
standard gravity which provide for $m_K=3$ directly the
well-known nonabelian chiral anomalies.
Nevertheless, every $q_K$ with $m_K>2$ has a local
$\4s$-invariant completion. This follows immediately from
the results of section \ref{susy}. Indeed, as $f_K$ is (a)
$\4s$-closed due to $f_K=\4sq_K$ and $\4s^2=0$, (b)
depends only on the $\4\xi$ and $\cT$, and (c)
has total degree $2m_K$, (\ref{c3}) implies
\beq m_K>2:\quad f_K=\4s\, p_K(\4\xi,\cT)\label{cons7a}\eeq
for some local total form $p_K(\4\xi,\cT)$.
Hence, the total forms
\beq \4q_K=q_K-p_K\label{cons8a}\eeq
are $\4s$-invariant,
\beq \4s\, \4q_K=0\quad (m_K>2).\label{cons9a}\eeq
Note that $\4q_K$ does not vanish as $p_K$ depends only
on the $\4\xi$ and $\cT$ whereas $q_K$
involves the $\4C$ too. Any polynomial $P(\Th)$
satisfying (\ref{comp3a}) can thus indeed be completed to an
$\4s$-invariant total form $P(\4q)$ by replacing
$\Th_K$ with the corresponding $\4q_K$. In particular
the $\4q_K$ with $m_K=3$, given explicitly in \cite{phd,sugra},
provide the supersymmetrized
versions of the nonabelian chiral anomalies spelled out
in section \ref{anos}.

Similar arguments prove that the remaining terms in
(\ref{comp3}) can be completed to local $\4s$-invariants. This
will be now shown for the second term in (\ref{comp3}) (the third term
can be treated in a completely analogous way).
In a first step we complete $\cH\7P(\Th,\4Q)$ to
\[ \alpha'=\cH\7P(q,\4Q)\]
by replacing in $\7P$ all $\Th_K$ with the corresponding
$q_K$. Thanks to (\ref{comp5}) and $\cH^2=0$ (the latter holds
as $\cH$ is Grassmann odd), the $\4s$-transformation of
$\alpha'$ reads
\[ \4s\alpha'=
-\cH f_K\, \frac{\6\7P(q,\4Q)}{\6q_K}\ .
\]
The total forms $\cH f_K$ occurring
on the r.h.s. of this equation
are $\4s$-closed, depend only on the
$\4\xi$ and $\cT$ and have total degrees $2m_K+3>4$.
(\ref{c3}) therefore implies the existence of local
total forms $h_K(\4\xi,\cT)$ such that
\[ \cH f_K=\4s\, h_K(\4\xi,\cT). \]
In a second step we now consider
\[ \alpha''=\alpha'+h_K\, \frac{\6\7P(q,\4Q)}{\6q_K}\ . \]
Its $\4s$-transformation is given by
\[ \4s\alpha''=h_{[K}f_{L]}\, \frac{\6^2\7P(q,\4Q)}{\6q_L\6q_K}
+h_K\cH\, \frac{\6^2\7P(q,\4Q)}{\6\4Q\6q_K}
\]
where the antisymmetrization in $K$ and $L$ in the first term
on the r.h.s. is automatic thanks to the
odd Grassmann parity of the $q_K$. $h_{[K}f_{L]}$ is $\4s$-closed
due to $\4s(h_{[K}f_{L]})=\cH f_{[K}f_{L]}=0$ (one has $f_{[K}f_{L]}=0$ 
because the $f_K$ are Grassmann even), depends only on the 
$\4\xi$ and $\cT$, and
has total degree $2+2m_K+2m_L>5$.
(\ref{c3}) therefore implies
\[ h_{[K}f_{L]}=-\4s h_{KL}(\4\xi,\cT)\]
for some $h_{KL}=-h_{LK}$. Analogous arguments
(using $\cH^2=0$ again) imply
\[ h_K\cH=-\4s\, g_K(\4\xi,\cT)\]
for some $g_K$. We conclude that the $\4s$-transformation of
\[ \alpha'''=\alpha''+
h_{KL}\, \frac{\6^2\7P(q,\4Q)}{\6q_L\6q_K}
+g_K\, \frac{\6^2\7P(q,\4Q)}{\6\4Q\6q_K}
\]
contains only third order derivatives of $\7P(q,\4Q)$
w.r.t.\ $\4Q$ and the $q_K$. 
The arguments can be iterated until $\4Q$ and all
the $q_K$ are differentiated away
and one is left with an $\4s$-invariant
completion $\alpha'{}^{\cdots}{}'$ of
$\cH \7P(\Th,\4Q)$.

Those $P(\Th,\4Q)$ which involve $\4Q$ or one of the
$\Th_K$ with $m_K=1,2$ and the terms in (\ref{comp1})
containing the $\cF^{\, i_a}$ do not provide further 
solutions to $\4s\alpha(\cW)=0$. Namely either 
they cannot be completed to $\4s$-invariants or
the respective $\4s$-invariants 
are cohomologically equivalent to solutions arising already
from (\ref{comp3}).
This can be proved as the analogous statement
in standard (non-supersymmetric) gravity,
cf. \cite{grav} for details.

To summarize, an $\4s$-invariant local completion exists for any
total form (\ref{comp3}) satisfying (\ref{comp3a}) and
these completions provide all the solutions
to $\4s\alpha(\cW)=0$ up to trivial ones of the
form $\4s\beta(\cW)$. The resulting list of solutions
is still overcomplete because it still contains trivial
solutions which may be removed at each total degree
separately. The solutions with total degree 4 and 5 have been
given explicitly in \cite{sugra} and will be
discussed in sections \ref{actions} and \ref{anos}. They
provide the supergravity Lagrangians and
the antifield independent candidate gauge anomalies respectively.
The other solutions (with higher total degrees)
will not be further discussed here because a physical interpretation
is not yet known for them (they can be found in \cite{phd}).

\mysection{Invariant Actions}\label{actions}

\subsection{Old minimal supergravity}\label{old}

Supergravity actions which are invariant under the
standard gauge resp.\ BRST transformations given in section
\ref{brs} arise from those total forms
(\ref{comp3}) which have total degree 4. In old minimal
supergravity these are just linear combinations of the
$\cP_\Delta$ (with constant coefficients)
because the first term in (\ref{comp3}) provides
only solutions with total degrees exceeding 4 due to
(\ref{comp3a}) and the second term contributes only in
new minimal supergravity.
Hence, in old minimal supergravity
the integrand of the most general real action $\int \omega_4$
that is invariant under the standard gauge transformations
is the 4-form $\omega_4$ contained in
$\cP=a^\Delta \cP_\Delta$ where $a^\Delta$ are
arbitrary real constant coefficients. 
It is an easy exercise to verify that the result is
\[ \omega_4=d^4x\, e\, L_{old}\ ,\quad e=\det(\viel a\mu),\]
with
\bea L_{old}&=& (\5\cD^2-4i\psi_\mu\sigma^\mu\5\cD-3M
+16\psi_\mu\sigma^{\mu\nu}\psi_\nu)\, \Omega+c.c.\ ,\nonumber\\
\Omega&=&\cA(\5M,\5W,\5\lambda)+(\cD^2-\5M)\, \cB(\cT)
\label{act2}\eea
where $\5\cD^2$ and
$\cD^2$ are shorthand notations for $\5\cD_\da\5\cD^\da$ and
$\cD^\alpha\cD_\alpha$ respectively,
$\cB\equiv a^\Delta \cB_\Delta$ is $\cG$-invariant whereas
$\cA\equiv a^\Delta \cA_\Delta$ is $\cG$-invariant 
except under $R$-transformations if the
latter are to be gauged (then $\cA$ must have $R$-charge $-2$).
It is of course well-known that supergravity actions can be 
constructed from (\ref{act2}). For instance, (\ref{act2})
emerges from Eq. (15.28) of \cite{des}
when one identifies the function $L$ occurring there 
with $\5\Omega$ given above, and can also be obtained
from superspace integrals \`a la \cite{wessbagger}. 
The new result we have derived here is 
that (\ref{act2}) gives remarkably the {\em most general} local
action for old minimal supergravity.

The supersymmetrized Einstein--Hilbert action arises
from a contribution to $\cA$ proportional to $\5M$ yielding
\bea M_{Pl}^{-2}L_{grav}&=&\s0 12\cR+2D^\dR
-2i\psi_\mu\sigma^\mu(\5S+i\5\lambda^\dR)
+2i(S-i\lambda^\dR)\sigma^\mu\5\psi_\mu\nonumber\\
& &-3B_aB^a-\s0 3{16} M\5M+
\s0 32(\5M\psi_\mu\sigma^{\mu\nu}\psi_\nu+
M\5\psi_\mu\5\sigma^{\mu\nu}\5\psi_\nu).
\label{act3}\eea
where $\lambda^\dR$ and $D^\dR$ contribute
of course only if $R$-transformations
are to be gauged (otherwise these fields 
simply have to be set to zero) and $M_{Pl}$ is the Planck mass.
I note that (\ref{act3}) is an unusual way to
write $L_{grav}$ but agrees in fact completely
with more familiar expressions that can be found in the
literature. For instance,
in (\ref{act3}) the super-covariantized curvature scalar
$\cR$ contains gravitino dependent contributions
that combine with the term
$2iS\sigma^\mu\5\psi_\mu+c.c.$ to the
familiar kinetic term for the gravitino given already in
\cite{invention}. Furthermore, all the terms linear in $B$, $M$
and $\5M$, i.e. those contained in $\cR$, $S$ and $\5S$ and
the last two terms in (\ref{act3}), cancel exactly.

The supersymmetrized Yang--Mills Lagrangian arises from
the contribution $\s0 1{16}\5\lambda^i\5\lambda_i$ to $\cA$
(nonabelian indices $i$ are lowered with
the Cartan--Killing metric of the Yang--Mills gauge group and
abelian ones with the unit matrix). It reads
\bea
L_{YM}=-\s0 14\, {F_{\mu\nu}}^iF^{\mu\nu}{}_i
-\s0 12\, i \, (\lambda^i\sigma^\mu\nabla_\mu\5\lambda_i
+\5\lambda^i\5\sigma^\mu\nabla_\mu\lambda_i)+\s0 12\, D^i D_i
+\s0 32\, \lambda^i\sigma^\mu\5\lambda_i B_\mu\nonumber\\
-\s0 12\, \F \mu\nu{i}\ep^{\mu\nu\rho\sigma}
(\psi_\rho\sigma_\sigma\5\lambda_i+\lambda_i\sigma_\sigma\5\psi_\rho)
+\psi_\mu\sigma^{\mu\nu}\psi_\nu\5\lambda^i\5\lambda_i
+\5\psi_\mu\5\sigma^{\mu\nu}\5\psi_\nu \lambda^i \lambda_i
\label{act4}\eea
where $\ep^{\mu\nu\rho\sigma}=\Viel a\mu\cdots\Viel d\sigma
\ep^{abcd}$ is vierbein dependent
($e\ep^{\mu\nu\rho\sigma}$ is constant), $\nabla_\mu $ is
the usual covariant derivative (not the super-covariant one),
\beq \nabla_\mu=\6_\mu-\A i\mu\delta_i-\s0 12\,\spin {ab}\mu l_{ab}\ ,
\label{hatcov}\eeq
and ${F_{\mu\nu}}^i$ is the super-covariantized Yang--Mills
field strength,
\beq {F_{\mu\nu}}^i=\6_\mu \A i\nu-\6_\nu \A i\mu
     +\f jki\A j\mu\A k\nu
     +2i\, (\lambda^i\sigma_{[\mu}\5\psi_{\nu]}
               +\psi_{[\mu}\sigma_{\nu]}\5\lambda^i).
\label{superF}\eeq

Of course (\ref{act2}) can be used to construct
supergravity actions that generalize the simple one arising
from (\ref{act3}) and (\ref{act4}). In particular, a
constant contribution $m$ to $\cA$ gives rise to
\beq L_{cosmo}= -3mM
+16m\psi_\mu\sigma^{\mu\nu}\psi_\nu+c.c.
\label{act5}\eeq
which, when included in $L_{old}$,
contributes to the cosmological constant.
Note however that $L_{cosmo}$ is neither locally nor
globally $R$-invariant and is thus forbidden when
global or local $R$-invariance is imposed, in contrast
to $L_{grav}$ and $L_{YM}$.
Note also that the most general
action contains at most one Fayet--Iliopoulos
contribution, namely the one for $R$-transformations
occurring in $L_{grav}$.

\subsection{New minimal supergravity}\label{new}

In new minimal supergravity, both the second and third
term in (\ref{comp3}) contain total forms with total degree 4
and thus give contributions to the most general invariant
action. The corresponding contributions to the Lagrangian are
 denoted by $L_1$ and $L_2$ respectively,
\beq L_{new}=L_1+L_2\ .\label{act6}\eeq
$L_1$ arises from the second term in
(\ref{comp3}) by choosing $\7P=\mu_{i_a}\4C^{i_a}$
as a linear combination of the abelian $\4C$'s with
constant coefficients $\mu_{i_a}$
(the $\4C^{i_a}$ are those $\Th_K$ with $m_K=1$).
Now, $\cH\7P$ is not yet $\4s$-invariant. We know however
that it can be completed to an $\4s$-invariant local total
form, see section \ref{completion}.
This completion has been computed already in \cite{sugra} and
is given by
\beq \mu_{i_a}\, (2\4C^{i_a}\cH
+\5\lambda^{i_a}\5\eta+\eta\lambda^{i_a}+\Xi \, D^{i_a})
\label{act8}\eeq
where $\Xi$ is the ``total volume form" (\ref{cp14}),
$\eta^\alpha$ is given by
\beq \eta^\alpha=-\s0 i6\, \4\xi_\dbe\4\xi^{\dbe\beta}
\4\xi_{\beta\da}\4\xi^{\da\alpha} \label{act9}\eeq
and it is understood that, in accordance with
appendix \ref{appA}, the gaugino and $D$-field of
$R$-transformations are identified with
\beq \lambda_\alpha^\dR\equiv -iS_\alpha\ ,\quad
D^\dR\equiv-\s0 14\, (\cR+H_{abc}H^{abc}).
\label{act10}\eeq
The appearance of the $D$-fields in (\ref{act8}) indicates
already that $L_1$ contains Fayet--Iliopoulos terms for the
abelian symmetries except for the
$R$-symmetry. The latter exception is due to (\ref{act10})
which also shows that the terms in (\ref{act8}) corresponding
to $R$-transformations provide the supersymmetrized
Einstein--Hilbert action in this case. 
With $ M_{Pl}^2=-\mu_\dR/2$, one obtains
from the volume form contained in (\ref{act8})%
\footnote{I remark that
(\ref{act12}) and (\ref{act13}) correct a mistake
in formula (3.29) of \cite{sugra}.}
\bea L_1&=&L_{grav}+L_{FI} ,\label{act11}\\
 M_{Pl}^{-2} L_{grav}&=& \s0 12\cR+\s0 12H_{abc}H^{abc}
+2i(S\sigma^\mu\5\psi_\mu-\psi_\mu\sigma^\mu\5S)
-2\ep^{\mu\nu\rho\sigma}\A \dR\mu \6_\nu t_{\rho\sigma} ,\ 
\label{act12}\\
L_{FI}&=&\sum_{i_a\neq \dR}
\mu_{i_a}  (D^{i_a}
+\lambda^{i_a}\sigma^\mu\5\psi_\mu+\psi_\mu\sigma^\mu\5\lambda^{i_a}
+\ep^{\mu\nu\rho\sigma}\A {i_a}\mu \6_\nu t_{\rho\sigma}).
\label{act13}\eea
Of course, by an appropriate choice of basis for the
abelian gauge multiplets one can assume that at most
one $\mu_{i_a}$ ($i_a\neq\dR$) is different from zero.
Again, one may check that
(\ref{act12}) agrees indeed completely with the
action for new minimal supergravity that can be found in the
literature, see e.g. \cite{new,mueller}.
(\ref{act12}) and (\ref{act13}) cannot naturally be written 
as standard superspace integrals, unless one modifies the whole
approach using an enlarged field content, 
c.f.\ \cite{bible,mueller}.

The remaining terms $L_2$ in the general Lagrangian
(\ref{act6}) are analogous to the Lagrangian
(\ref{act2}) of old minimal supergravity and can thus
be written as superspace integrals. The difference to (\ref{act6}) 
is of course that the auxiliary field $M$ is absent now. Therefore
the supersymmetrized Einstein--Hilbert action does not arise from
$L_2$ which reads
\bea L_2&=&(\5\cD^2-4i\psi_\mu\sigma^\mu\5\cD
+16\psi_\mu\sigma^{\mu\nu}\psi_\nu)\, \Omega+c.c.\ ,\nonumber\\
\Omega&=&\cA(\5W,\5\lambda)+\cD^2 \cB(\cT)
\label{act14}\eea
where $S$ counts among the $\lambda$'s due to (\ref{act10}).
The supersymmetrized Yang--Mills action arises from (\ref{act14})
through a contribution proportional to
$\sum_{i\neq\dR}\5\lambda^i\5\lambda_i$ to $\cA$,
like in old minimal supergravity (the sum over $i$
excludes $\dR$ here in order to end up with the standard action for new
minimal supergravity). Since $R$-transformations
are gauged in new minimal supergravity, $\cA$ must have 
$R$-charge $-2$ and therefore does not contain a
constant piece in this case. Hence, $L_{new}$
contains no contribution analogous to (\ref{act5}).

\mysection{Full BRST cohomology}\label{full}

We will now analyze the full BRST cohomology
based on the standard super-Einstein--Yang--Mills action
$\int d^4x\, e\, L$ with Lagrangian
\beq L=L_{grav}+L_{YM}\label{f1}\eeq
where $L_{grav}$ is given for old and new minimal
supergravity by (\ref{act3}) and (\ref{act12}) respectively,
and $L_{YM}$ as in (\ref{act4}). The specialization to a
particular action is necessary because the BRST transformations
of the antifields involve the variational derivatives
of the action with respect to the fields.
Choosing the simple action with
Lagrangian (\ref{f1}) has the major advantage that the results can be
easily generalized to more complicated actions by means of spectral
sequence techniques, cf. remarks in section \ref{matter}.

The analysis of the full BRST cohomology proceeds
closely to that of the restricted BRST cohomology in
section \ref{rest}. It uses the fact that the full BRST cohomology
(taking the antifields into account) can be obtained from the
weak (``on-shell") $\4s$-cohomology involving the fields only
(but not the antifields) \cite{ten}.

\subsection{On-shell basis for the tensor fields}\label{onbasis}

It will be crucial to determine an appropriate
{\em on-shell basis} for the tensor fields. Such a basis
is a subset of the off-shell basis
determined in section \ref{basis} taking the
equations of motion into account.
This makes sense because the equations of motion are
equivalent to equations involving only the tensor fields 
\cite{ten}. The equations of motion can therefore be used to
express some of the tensor fields $\cT$ in terms
of others (of course they may even set some $\cT$'s
to zero). The remaining $\cT$'s form the searched for
on-shell basis for the tensor fields which will be denoted
by $\{\7\cT^\TIX\}$.

Let me first illustrate the procedure for
pure old minimal supergravity with Lagrangian (\ref{act3}) 
without gauged $R$-symmetry (i.e. for
$\lambda^\dR=D^\dR=0$). In this case the equations of
motion simply set
$\cR$, $Y$, $S$, $U$, $M$ and $B$ to zero where the notation
of section \ref{basis} is used and spinor indices are suppressed%
\footnote{This is a
somewhat unusual but correct form of the equations of motion.
E.g., the gravitino dependent terms that appear in more familiar
versions of the equations of motion
``on the r.h.s.\ of the Einstein equations" are indeed taken into
account as $\cR$ and $Y$ are super-covariant.}. Using
(\ref{bas13}), one now easily verifies that an on-shell basis
for the tensor fields in pure old minimal supergravity
is given by $W$, $\5W$, $X$, $\5X$ and all their $\cD^+_+$
derivatives defined in (\ref{bas4})
(recall that $X$ is the super-covariantized
Weyl tensor and that $W$ is the chiral part of the
super-covariantized gravitino field strength).

When Yang--Mills multiplets
are present,  $\cR$, $Y$, etc.
do not vanish anymore on-shell but can still be expressed
(nonlinearly) in terms of other tensor fields.
In addition
one has the equations of motion for the Yang--Mills multiplets which
are analyzed analogously. One obtains that
an on-shell basis for the tensor fields is given by
\beq
\{\7\cT^\TIX\}=\{W_q,\, \5W_q,\, X_q,\, \5X_q,\, \lambda^i_q,\,
\5\lambda^i_q,\, G^i_q,\, \5G^i_q:\ q=0,1,\ldots\}
\label{f3}\eeq
where $q$ indicates the number of $\cD^+_+$ operations,
\beq W_q\equiv (\cD^+_+)^q\, W\quad etc.
\label{f3a}\eeq
In new minimal supergravity one finds analogously that
(\ref{f3}) gives
an on-shell basis for the tensor fields with the understanding
that it does not contain any tensor field associated with
local $R$-symmetry ($G^\dR$ and $\5G^\dR$ are eliminated
using the equations of motion for $t_{\mu\nu}$ while $H_{abc}$
is eliminated through the equations of motion for $\A \dR\mu$).

\subsection{Sketch of the computation}\label{step1}

As shown in \cite{ten}, the computation of the
full BRST cohomology reduces (locally) to the determination
of the weak cohomology of $\4s$ on local total forms depending
only on the $\cW$'s given in (\ref{Ws}). The computation of
this cohomology can therefore be further reduced to a problem
involving only the tensor fields
$\7\cT$ given in (\ref{f3})
and the generalized connections $\4\xi^A$, $\4C^I$ and
$\4Q$ where it is understood
that $G^\dR$ and $\5G^\dR$ do not count among the $\7\cT$
in new minimal supergravity and that $\4Q$ is absent in
old minimal supergravity. To formulate the problem on
the remaining variables correctly, one has
to express their weak $\4s$-transformations
completely in terms of these variables again. To that end one must
use the equations of motion to express those
tensor fields $\cT\not\in\{\7\cT^\TIX\}$
which occur in $\4s\7\cT$, $\4s\4\xi$, $\4s\4C$ and $\4s\4Q$
in terms of
the $\7\cT$ as described in section \ref{onbasis}.
Denoting weak (= on-shell) equalities by $\approx$,
we are thus left with the problem
\beq \4s\, \alpha(\4C,\4\xi,\4Q,\7\cT)\approx 0
\label{f4}\eeq
defined modulo trivial solutions which are weakly of the
form $\4s\beta(\4C,\4\xi,\4Q,\7\cT)+constant$.

(\ref{f4}) is now analyzed analogously to (\ref{r2}).
As the equations of motion do not interfere
with the Lie algebra cohomology, one first shows
as in section \ref{decomp} that
the nontrivial solutions of (\ref{f4}) are at
highest degree in the $\4C$ and $\4Q$ given by
\beq f_i(\4\xi,\7\cT)\, P^i(\Th,\4Q)
\label{f5}\eeq
where the $f_i$ solve
\beq \4s\, f_i(\4\xi,\7\cT)\approx 0,\quad
f_i(\4\xi,\7\cT)\not\approx \4s\, h_i(\4\xi,\7\cT).
\label{f6}\eeq

The problem (\ref{f6}) is now analyzed using techniques
similar to those described in section \ref{susy}.
There is however one important complication compared
to the restricted cohomology. It consists in the
fact that (\ref{f4}) has in general
nontrivial solutions with total degree $G<4$, in contrast
to the analogous ``strong" problem (\ref{r12}). Such solutions
correspond to local conservation laws
as they provide representatives of the local
BRST cohomology at negative ghost numbers \cite{bbh1}. 
For $G<3$ they can be computed using methods developed in
\cite{bbh1} (see also \cite{HKS}) which are not repeated
here. One finds that there are no
nontrivial solutions with total degrees $G<1$, whereas the
only nontrivial solutions at $G=1$ and $G=2$ correspond to
solutions of (\ref{i1}) given by
\bea G=1&\leftrightarrow &
d^4x\, Q^*,\label{f8}\\
G=2& \leftrightarrow&
d^4x\, C^*_{i_a}\quad (i_a\neq\dR),\label{f7}\eea
where $Q^*$ and $C^*_{i_a}$ are the antifields of
$Q$ and $C^{i_a}$ respectively.
It is very easy to verify that (\ref{f8}) and (\ref{f7})
solve (\ref{i1}). Indeed (\ref{ss0}) gives directly
\beq sQ^*=\6_\mu (C^\mu Q^*+Q^{\mu *});\quad i_a\neq\dR:\ 
s C^*_{i_a}=\6_\mu (C^\mu C^*_{i_a}-A^{\mu*}_{i_a}).
\label{zusatz}\eeq
Of course, (\ref{f8})
occurs only in new minimal supergravity. There is no
solution (\ref{f7}) corresponding to $R$-transformations
because the gravitino and the gauginos have non-vanishing
$R$-charges.
Analogously $d^4x C^*_{i_a}$ disappears from the list of solutions
when one includes further (matter) fields transforming nontrivially
under the $i_a$th abelian gauge symmetry, see \cite{bbh1}.

(\ref{zusatz}) ensures that (\ref{f8}) and (\ref{f7}) give rise to
solutions of $\4s\alpha=0$. The latter are obtained by evaluating 
the descent equations implied by (\ref{zusatz}).
We denote these solutions by $\4Q^*$ and
$\4C^*_{i_a}$ respectively. It is easy to verify that
they are of the form
\bea
\4Q^*&=&\s0 1e\,\Xi\, Q^*
-\s0 1{6e}\, \4\xi^a\4\xi^b\4\xi^c\ep_{abcd} Q^{d*}
+\dots+4\4C^\dR,\label{f10}\\
\4C^*_{i_a}&=&\s0 1e\,\Xi\,C^*_{i_a}
+\s0 1{6e}\, \4\xi^a\4\xi^b\4\xi^c\ep_{abcd} A_{i_a}^{d*}
+\dots \quad(i_a\neq\dR)
\label{f10a}\eea
where $\Xi$ is the total volume form (\ref{cp14}).
Furthermore one finds that the antifield independent part
of (\ref{f10a}) is given by
\beq \4C^*_{i_a|}=\s0 14\, \4\xi^a\4\xi^b\ep_{abcd}F^{cd}_{i_a}
+\th\lambda_{i_a}-\5\th\5\lambda_{i_a}\label{newsol}\eeq
with
\beq\th^\alpha=\4\xi_\da\4\xi^{\da\alpha},
\quad \5\th^\da=\4\xi^{\da\alpha}\4\xi_\alpha\ .\eeq
(\ref{newsol}) thus solves (\ref{f6}) as it
depends only on the $\4\xi$ and $\7\cT$.
In contrast, $\4Q^*$ involves $\4C^\dR$
and therefore its antifield independent part
solves (\ref{f4}) but {\em not} (\ref{f6}).
The latter reflects the presence of the
Chern--Simons like term
$\ep^{\mu\nu\rho\sigma}\A \dR\mu \6_\nu t_{\rho\sigma}$
in (\ref{act12}). The antifield independent parts
of the $\4C^*_{i_a}$ are therefore the only
nontrivial solutions to (\ref{f6})
with total degree $G<3$ on top of those occurring
already in (\ref{c3}).
The latter remain indeed nontrivial even
in the weak $\4s$-cohomology on local total forms
$f(\4\xi,\7\cT)$, except for $\cF^\dR$ in the
case of new minimal supergravity where it vanishes weakly. 
This quite plausible statement can
be proved rigorously by a technique used in
appendix \ref{CPL}. The proof parallels that of
a corresponding result in standard gravity
given in appendix E of \cite{bbhgrav}
and is therefore not spelled out here.

Let us now
turn to the discussion of (\ref{f6}) for
total degrees $G\geq 3$. Similarly to the analogous
``strong" problem for $G\geq 4$ in section \ref{susy}
we decompose (\ref{f6})
according to the degree in the tensor fields $\7\cT$.
At lowest degree, this yields the linearized problem
\beq \delta_{susy} f_{\tdeg \ell}(\4\xi,\7\cT)\sim 0,
\quad f_{\tdeg \ell}\not\sim
\delta_{susy} h_{\tdeg \ell}(\4\xi,\7\cT),
\quad \delta_If_{\tdeg \ell}=\delta_Ih_{\tdeg \ell}=0
\label{f11}\eeq
with $\delta_{susy}$ as in (\ref{c8})--(\ref{c9b}), and $\sim$
denoting ``linearized weak equality" based on the
linearization of the equations of motion 
in the tensor fields $\7\cT$.

Now, in the cases $G\geq 4$ the methods and results
of \cite{glusy}
can be straightforwardly adapted to solve (\ref{f11}).
Again, this is possible thanks to the
QDS structure of the on-shell representation
of the subalgebra $\{D_\alpha,D_\beta\}=0$ of (\ref{c10}) proved in
appendix \ref{onqds}.
This implies that, in the cases $G\geq 4$, the
solution of (\ref{f6}) is analogous to the solution of the
corresponding ``strong" problem (\ref{c99}), 
see appendix \ref{global} for details.

One is left with the case $G=3$
which I have not been able to solve completely. Partial
results are derived in appendix \ref{global} where it is shown
that all the solutions to (\ref{f11}) with $G=3$
can be assumed to be of the form
\bea G=3:\quad f_{\tdeg \ell}=\{4\Theta
-i\5\th_\da\4\xi^{\da\alpha}D_\alpha
-i\th^\alpha\4\xi_{\alpha\da}\5D^\da
\nonumber\\
+\s0 1{12}\4\xi^{\da\beta}\4\xi_{\beta\dbe}\4\xi^{\dbe\alpha}
[D_\alpha,\5D_\da]\}\,  R(\7\cT)\label{f12a}\eea
where $\Theta$ is the quantity
\beq \Theta=\4\xi^\alpha\4\xi_{\alpha\da}\4\xi^\da
\label{f12}\eeq
and $R(\7\cT)$ are real functions solving
\beq D^\alpha D_\alpha R(\7\cT)\sim 0,\quad R=\5R.\label{f12b}\eeq
One solution which is always present is of course
$R=constant$ for which (\ref{f12a}) becomes simply proportional to
$\Theta$. The latter can be completed to a solution of (\ref{f6})
corresponding to the Noether current for $R$-transformations in the
case of old minimal supergravity
(see section \ref{noether}) and reproducing 
in new minimal supergravity the solution
$\cH$ already present in the restricted cohomology, cf. (\ref{c3}),
where now of course $\cH$ has to be replaced by its on-shell
version (e.g. in pure new minimal supergravity
$\cH$ reduces on-shell to $i\Theta$).
Recall also that $\cH$ is $\4s$-exact, cf. (\ref{s8}), but it
is not $\4s$-exact in the space of total forms $f(\4\xi,\7\cT)$
(not even weakly).

There might be further solutions to (\ref{f6})
with $G=3$, in particular when matter
multiplets are included. Fortunately it will not matter
in the following whether we know all these solutions
explicitly. We denote them by $N_\tau$, except for the
special solution $\cH$ present only in
new minimal supergravity which is treated separately because it is
the only one which occurs already in the restricted cohomology.
We can then summarize the solution of (\ref{f6}) as follows:
\bea
& &\4s\, f(\4\xi,\7\cT)\approx 0,\qd \tot(f)=G\nn\\
\LRA& &\qd f(\4\xi,\7\cT)\approx
\left\{\ba{ll}
constant & \mbox{for $G=0$}\\[.5ex]
\4sh(\7\cT) & \mbox{for $G=1$}\\[.5ex]
a_{i_a}\cF^{\,i_a}+\sum_{i_a\neq \dR}b^{i_a}\4C^*_{i_a|}
+\4sh(\4\xi,\7\cT) & \mbox{for $G=2$}\\[.5ex]
a\, \cH+a^\tau N_\tau(\4\xi,\7\cT)
+\4sh(\4\xi,\7\cT)& \mbox{for $G=3$}\\[.5ex]
a^\Delta \7\cP_\Delta+\4sh(\4\xi,\7\cT)& \mbox{for $G=4$}\\[.5ex]
\4sh(\4\xi,\7\cT)& \mbox{for $G>4$}
\ea\right.
\label{f13}\eea
where the $a$'s and the $b^{i_a}$ are constants
($a_\dR$ can be assumed to vanish in new minimal supergravity
as $\cF^\dR$ vanishes on-shell in that case, see above),
$\4C^*_{i_a|}$ was given in (\ref{newsol}), and
\beq
\cP_\Delta =\7\cD_\da \7\cD^\da\, \Xi\,
\{ \cA_\Delta(\5W,\5\lambda)+
\cD^\alpha \cD_\alpha \cB_\Delta(\7\cT)\}+c.c.
\label{f14}\eeq
with $\7\cD_\da$ as in (\ref{cp15b}).
In (\ref{f13}) it is understood that
the on-shell version of all occurring
total forms (esp. of $\cH$) is used.

\subsection{Result}\label{result}

We can now complete the analysis of the full BRST cohomology. 
To that end we have to find out which total forms
\beq
\7P+\cF^{\, i_a}\7P_{i_a}
+\sum_{i_a\neq\dR}\4C^*_{i_a}P^{i_a}
+ \cH P_\cH
+N_\tau P^\tau+
\7\cP_\Delta P^\Delta
\label{f15}\eeq
can be completed to (inequivalent) $\4s$-invariants, where
the $P$'s and $\7P$'s are polynomials in the $\Th_K$ and in $\4Q$.
Without going into details I note that arguments analogous
to those used in section \ref{completion} show that
\ben
\item $\7P$ depends neither on $\4Q$ nor on those $\Th_K$
with $m_K=1,2$ except on $\4C^\dR$ in new
minimal supergravity. The latter exception reflects that
$\4C^\dR$ can be completed to an $\4s$-invariant
total form in new minimal supergravity, cf.  (\ref{f10}). Hence, we get
$\7P=P(\Th)+\4C^\dR P_{new}(\Th)$  where
$P_{new}$ is present only in new minimal
supergravity and
\beq 
\frac {\6P(\Th)}{\6\Th_K}=\frac {\6P_{new}(\Th)}{\6\Th_K}=0\quad
\mbox{for}\quad m_K=1,2.
\label{f16}\eeq
\item The terms
$\cF^{\, i_a}\7P_{i_a}$ either cannot be completed
to $\4s$-invariants or they can be removed by subtracting $\4s$-exact
total forms and redefining $P_\cH$ and $P^\Delta$ appropriately.
\item The terms $\4C^*_{i_a}P^{i_a}$ ($i_a\neq\dR$) must
be of the  form
\beq 
\sum_{i_a\neq\dR}
\4C^*_{i_a}\, \frac{\6P_*(\Th,\4Q)}{\6\4C^{i_a}}
\label{f17}\eeq
where $P_*$ is an arbitrary polynomial
in the case of new minimal supergravity, whereas
in old minimal supergravity it must not involve $\4C^\dR$,
\beq \frac{\6P_*(\Th,\4Q)}{\6\4C^\dR}=0\quad
\mbox{in old min. supergravity.}
\label{f18}\eeq
\een
The remaining total forms (\ref{f15}), summarized in
table 6.1, have $\4s$-invariant
completions which provide (locally) a complete set of
cohomology classes of the full $\4s$-cohomology.
This set is actually still overcomplete because
it still contains trivial (i.e.\ $\4s$-exact) total forms.
The latter may be removed at each total degree
separately. The results for the physically
important cases (total degrees $\leq 5$)
are spelled out and discussed
in the following sections. Note that
the solutions of type III and VI are present only in new minimal
supergravity and that the solutions I--III can be
chosen so as not to involve antifields when the auxiliary fields
are used (as they have
counterparts in the restricted cohomology, see section
\ref{completion}). The remaining solutions
IV--VI necessarily involve antifields, whether or not
the auxiliary fields are used. 
\[
\ba{c|l|l}
\mbox{Type} & \alpha & \mbox{Remarks}\\    
\hline\rule{0em}{3ex}
\mbox{I} & P(\Th)+\dots &
    \mbox{$P$ as in (\ref{f16})}\\
\rule{0em}{3ex}
\mbox{II} & \7\cP_\Delta P^\Delta(\Th,\4Q)+\dots & \\
\rule{0em}{3ex}
\mbox{III} & \cH P_\cH(\Th,\4Q)+\dots & \\
\rule{0em}{3ex}
\mbox{IV} & N_\tau P^\tau(\Th,\4Q)+\dots& \\
\rule{0em}{3ex}
\mbox{V} &\sum_{i_a\neq\dR}
\4C^*_{i_a}\, \6P_*(\Th,\4Q)/\6\4C^{i_a}+\dots &
\mbox{$P_*$ as in (\ref{f18})}\\
\rule{0em}{3ex}
\mbox{VI} & \4Q^* P_{new}(\Th)+\dots &
    \mbox{$P_{new}$ as in (\ref{f16})}\\
\multicolumn{3}{c}{}\\
\multicolumn{3}{c}{\mbox{Table 6.1:
$\4s$-cohomology}}
\ea
\]

\mysection{Dynamical conservation laws}
\label{noether}

The local conservation laws are determined by the {\em weak
cohomology of $d$} on
local differential forms at form degrees
$0<p<4$ (as the spacetime dimension is 4 and constant
zero forms are not counted among the
local conservation laws). In other words, they are the
solutions of
\beq d\, j_{4-k}\approx 0,\quad j_{4-k}\not\approx dk_{4-k-1}
\quad (k=1,2,3)\label{n1}\eeq
or, equivalently, of
\beq \6_{\mu_1}j^{\, [\mu_1\cdots\mu_k]}\approx 0,
\quad j^{\, \mu_1\cdots\mu_k}\not\approx
\6_{\mu_0}k^{ [\mu_0\cdots\mu_k]}
\label{n1a}\eeq
where the $j^{\,\mu_1\cdots\mu_k}$ are local functions of the 
classical fields and the $j_{4-k}$ are local $(4-k)$-forms
\[ j_{4-k}=\frac{1}{(4-k)!}\, dx^{\mu_4}\cdots dx^{\mu_{k+1}}
\, \frac 1e\,\ep_{\mu_4\cdots\mu_1}
j^{\,\mu_k\cdots\mu_1}.\]
We call $j_{4-k}$ a ``dynamical conservation law of order $k$"
when it solves (\ref{n1}) {\em locally}. Weakly $d$-closed
forms which are locally, but not globally $d$-exact on-shell
are called ``topological conservation laws" instead and
are briefly discussed in
section \ref{top}. The dynamical conservation laws of order 1
are the Noether currents $j^\mu$ and thus correspond
one-to-one to the nontrivial global symmetries of the 
classical action%
\footnote{A global symmetry $\delta_\epsilon\phi^i$
is called trivial
if it equals a special gauge transformation
(with special, possibly field dependent `parameter')
up to an on-shell vanishing
part of the form $\mu^{ij}\delta {\cal S}_{cl}/\delta\phi^j$
with $\mu^{ij}=-(-)^{\varepsilon_i\varepsilon_j}\mu^{ji}$
(in de Witt's notation).
Trivial global symmetries
correspond to trivial
Noether currents (satisfying $j^\mu\approx
\6_\nu k^{[\nu\mu]})$ and vice versa \cite{bbh1}.}.

The weak $d$-cohomology at form degree $(4-k)$
can be shown to be (locally) isomorphic
to the local BRST-cohomology at
{\em negative} ghost number $(-k)$ \cite{bbh1}.
The dynamical conservation laws
of order 1,2,3 are thus obtained from
the (full) $\4s$-cohomology
at total degree 3,2,1 respectively. The latter is obtained
from table 6.1 in section \ref{result}.
$j_p$ is just the antifield independent
part of the $p$-form contained in the corresponding
$\4s$-invariant total form $\alpha$ with total degree $p$.

The results are summarized in table 7.1 where $\4Q^*$ is the
$\4s$-invariant completion of $4\4C^\dR$ in new minimal
supergravity, cf. eq. (\ref{f10}), and
$j_\tau^\mu$ denotes the Noether current corresponding to
$N_\tau$. I note that table 6.1 contains one more
solution with total degree 3, namely the solution of type III
given just by $\cH$. The latter is however $\4s$-exact,
cf. (\ref{s8}), and does therefore not provide a 
dynamical conservation law.
\[
\ba{c|c|ll|l}
k & \mbox{Type}             &  \alpha & & j^{\, \mu_1\cdots\mu_k}\\
\hline\rule{0em}{3ex}
1 & \mbox{IV} & N_\tau && j^\mu_\tau\\
\rule{0em}{3ex}
 & \mbox{Va} & \4C^{*[i_a}\4C^{i_b]}+\ldots & (i_a,i_b\neq\dR)&
    e\, F^{\nu\mu[i_a}\A {i_b]}\nu+\ldots\\
\rule{0em}{3ex}
 & \mbox{Vb} & \4C^*_{i_a}\4Q^* & (i_a\neq\dR)&
    e\, F^{\nu\mu}_{i_a}\A \dR\nu+\ldots\\
\rule{0em}{3ex}
2 & \mbox{V} & \4C^*_{i_a} & (i_a\neq\dR)&
    \s0 12\, e\, F_{i_a}^{\mu\nu}+\ldots\\
\rule{0em}{3ex}
3 & \mbox{VI} & \4Q^* &&
\s0 23\, e\, \ep^{\mu\nu\rho\sigma}\A \dR\sigma+\ldots\\
\multicolumn{5}{c}{}\\
\multicolumn{5}{c}{\mbox{Table 7.1:
Dynamical conservation laws}}
\ea
\]
Let me briefly comment the result.

The third order conservation law 
occurs only in new minimal supergravity. In complete form
it reads
\beq j^{\mu\nu\rho}=e\, \ep^{\mu\nu\rho\sigma}
(\s0 23\, \A \dR\sigma
-\s0 14\, \lambda^i\sigma_\sigma\5\lambda_i)
+e\, H^{\mu\nu\rho}\label{j1}\eeq
with $H_{\mu\nu\rho}$ as in (\ref{identify}). 
This conservation law remains when new minimal supergravity
is coupled to matter fields in the standard way (then 
$j^{\mu\nu\rho}$ just
receives further terms involving the matter fields).
Old minimal supergravity cannot possess a
conservation law of third order as its gauge symmetries are
irreducible \cite{bbh1}.

The second order conservation laws read in complete form
\beq j^{\mu\nu}_{i_a}= \s0 12\, e\, F_{i_a}^{\mu\nu}
+ \s0 12\, e\,\ep^{\mu\nu\rho\sigma}
(\psi_\rho\sigma_\sigma\5\lambda_{i_a}
+\lambda_{i_a}\sigma_\sigma\5\psi_\rho),\quad i_a\neq \dR
\label{j2}\eeq
where ${F_{\mu\nu}}^{i_a}$ are the abelian super-covariant
field strengths (\ref{superF}).
These conservation laws disappear when matter fields are
included transforming nontrivially under the abelian gauge
transformations, cf. remarks in the text after (\ref{zusatz}).

The Noether currents associated with
the type-Va-solutions correspond to
the global symmetry of the Lagrangian (\ref{act4}) under
$SO(N)$ rotations of the $N$ abelian gauge multiplets different from
the one gauging $R$-transformations.

The type-Vb-solutions $\4C^*_{i_a}\4Q^*$
are present only in new minimal supergravity. 
They correspond to global symmetries
of new minimal supergravity transforming
for instance $t_{\mu\nu}$ among
others into the dual of the field strength of
$\A {i_a}\mu$ and $\A {i_a}\mu$ among others into a
linear combination of $\A \dR\mu$ and the dual of $H_{\mu\nu\rho}$ 
(this can be read off from the
antifield dependent terms contained in $\4C^*_{i_a}\4Q^*$).

The remaining Noether currents arise from the $N_\tau$.
Explicitly we know only one of them,
arising from the quantity $\Theta$ given in
(\ref{f12}) and present only in old minimal supergravity
when $R$-transformations are not gauged.
In {\em pure} old minimal supergravity $\Theta$ is
weakly $\4s$-invariant by itself. The 3-form contained in
it is given by
\[ dx^\mu dx^\nu dx^\rho\, \psi_\rho\sigma_\nu\5\psi_\mu\]
and provides the Noether current corresponding to the global
$R$-invariance of the Lagrangian (\ref{act3}) (with
$\lambda^\dR=0$ and $D^\dR=0$). When old minimal
supergravity is coupled to Yang--Mills multiplets via
(\ref{act4}), $\Theta$ is not weakly $\4s$-invariant anymore
by itself but
has still an $\4s$-invariant completion which again provides
the $R$-Noether current (containing now the gauginos as well)
and remains nontrivial unless $R$-transformations are gauged.

\mysection{On-shell counterterms and deformations}
\label{counter}

In this section we discuss the implications of the results
for the possible on-shell counterterms and for
the consistent and continuous deformations of
old and new supergravity and their gauge symmetries,
based on (\ref{f1}).
In fact both issues are closely related. 

The possible counterterms that are non-vanishing and gauge invariant
on-shell are directly determined by the local BRST cohomology at 
ghost number 0 as the latter is equivalent to the
weak BRST cohomology on local functionals constructed only out
of the classical fields.

That the consistent and continuous deformations are also restricted
by the local BRST cohomlogy at ghost number 0 was shown
in \cite{bh} where a systematic method was outlined to
obtain and classify these deformations.
The basic idea of this method is to deform the
(classical) master equation \cite{zj,bv}.
More precisely one looks for
a local solution to the master equation of the form
\beq \cS_g=\cS+g\,\cS^{(1)}+g^2\cS^{(2)}+\ldots
\label{co0}\eeq
where $\cS$ is the solution in the original (undeformed) theory
and $g$ is a deformation parameter. The master equation
$(\cS_g,\cS_g)=0$ is then expanded in powers of $g$
which yields
\beq \left(\cS,\cS^{(1)}\right)=0,\quad
\left(\cS^{(1)},\cS^{(1)}\right)+2\left(\cS,\cS^{(2)}\right)=0,
\quad \ldots
\label{co0a}\eeq
As $(\cS,\ \cdot\ )$ generates the original BRST transformations,
the first condition in (\ref{co0a}) requires
$\cS^{(1)}$ to be invariant under the undeformed BRST transformations.
To first order in $g$ the deformations
are thus indeed determined by the local BRST cohomology at ghost
number 0. At higher orders in $g$ the BRST cohomology at ghost
number 1 can impose further obstructions \cite{brandt}.
The antifield independent part of $\cS_g$ gives
of course the deformation of the original action while
the antifield dependent part provides the correspondingly deformed
gauge transformations. 

The BRST cohomology at ghost number 0 arises
from the $\4s$-cohomology at total degree 4.
The latter is obtained from section \ref{result}, leading to the
results summarized in table 8.1. The latter sketches both
the $\4s$-invariant total forms and the corresponding
BRST invariant local functionals, writing the latter as
$\int d^4x\, e\, L_{onshell}$ and giving only a
characteristic term. Moreover we use the notation
\beq \Omega_\Delta=\cA_\Delta(\5\lambda,\5W)+D^2\cB_\Delta(\7\cT),\quad
D^2=D^\alpha D_\alpha\, ,\quad \5D^2=\5D_\da \5D^\da\label{Omega}\eeq
with $\7\cT$ as in section \ref{onbasis} and
$D_\alpha \7\cT$ as in appendix \ref{onqds}, and
\beq J^\mu_\tau=\frac 1e\, j^\mu_\tau\label{covcurrents}\eeq
where $j^\mu_\tau$ is the Noether current corresponding
to $N_\tau$. $J^\mu_\tau$
is of course only covariantly conserved, in contrast to
$j^\mu_\tau$ which satisfies $\6_\mu j^\mu_\tau\approx 0$.
\[
\ba{c|ll|l}
\mbox{Type} &  \alpha\ &  & L_{onshell}\\ 
\hline\rule{0em}{3ex}
\mbox{II} & \7\cP_\Delta && 
\5D^2\Omega_\Delta+D^2\5\Omega_\Delta+\dots\\
\rule{0em}{3ex}
\mbox{III} & \cH \4C^{i_a}+\dots& &
    \A {i_a}\mu\6_{\nu}t_{\rho\sigma}\ep^{\mu\nu\rho\sigma}+\dots\\
\rule{0em}{3ex}
\mbox{IV} & \4C^{i_a}N_\tau+\dots &&
    \A {i_a}\mu J^\mu_\tau+\dots\\
\rule{0em}{3ex}
\mbox{Va} & \4C^{*[i_a}\4C^{i_b}\4C^{i_c]}+\dots& 
(i_a,i_b,i_c\neq\dR)&
   F^{\mu\nu [i_a}\A {i_b}\nu\A {i_c]}\mu+\dots\\
\rule{0em}{3ex}
\mbox{Vb} & \4C^{*[i_a}\4C^{i_b]}\4Q^*+\dots& (i_a,i_b\neq\dR)&
   4F^{\mu\nu [i_a}\A {i_b]}\nu\A \dR\mu+\dots\\
\rule{0em}{3ex}
\mbox{Vc} & \4C^*_{i_a}\4Q+\dots& (i_a\neq\dR) &
-\s0 12\, F^{\mu\nu}_{i_a}t_{\mu\nu}+\dots\\
\multicolumn{4}{c}{}\\
\multicolumn{4}{c}{\mbox{Table 8.1:
Counterterms and deformations}}
\ea
\]
Let me comment the result.

\underline{Type II and III}:
Using the auxiliary fields, all these functionals
can be completed to off-shell invariants which do not
involve antifields
as they have counterparts in the restricted BRST cohomology,
see section \ref{actions}.
The solutions of type II and III yield thus only
deformations which do not change the gauge transformations
nontrivially. I note that these solutions
yield among others
possible counterterms which were discussed already in
\cite{PvN}.
For instance, the integrand of 
a counterterm arising from
$\cB_\Delta= W^2\5W^2X^{2n}\5X^{2n}$
contains a contribution
proportional to $eX^{2(n+1)}\5X^{2(n+1)}$, i.e.
a term of order $4(n+1)$ in the Weyl tensor.
The terms of type III
are present only in new minimal supergravity
and reproduce the supergravity
Lagrangian (\ref{act12}) and the Fayet--Iliopoulos term
(\ref{act13}).

\underline{Type IV and V}:
The remaining terms have no counterparts in the restricted
$\4s$-cohomology. They yield therefore the
possible counterterms that are invariant on-shell but
cannot be completed to off-shell invariants, and provide
the possible nontrivial deformations of the gauge transformations
to first order in the deformation parameter.
We have to ask ourselves what these deformations
of the gauge symmetries might be.

The terms of type IV contain couplings of Noether currents 
to abelian gauge fields. This suggests that
the resulting deformations just gauge global symmetries in the
standard way.

The terms of type Va are reminiscent of the
trilinear vertex of gauge fields in nonabelian Yang--Mills theory
which suggests that the corresponding deformations
convert abelian gauge multiplets into
nonabelian ones. The complete antisymmetrization
in the group indices then corresponds to the antisymmetry
of the structure constants of the Lie algebra. The
Jacobi identity for these structure constants arises at
second order in $g$, cf. \cite{bht,brandt}
for a discussion in the nonsupersymmetric case.

The terms of type Vb occur only in new minimal
supergravity (due to (\ref{f18})) and
are somewhat similar to those of type Va. The differences
to the type-Va-terms are however that a) the antisymmetrization
in the abelian indices excludes the $R$-transformation and
b) the gauginos
and the gravitino transform nontrivially under $R$-transformations, 
in contrast to the properties of
the type-Va-terms. These differences reflect again
that $\4C^\dR$ can be completed to an $\4s$-invariant
total form in new minimal supergravity, cf. (\ref{f10}).
I have not yet investigated whether this might lead to
interesting unknown deformations.

The term of type Vc has no counterpart in standard Yang--Mills 
theory or gravity because it involves $t_{\mu\nu}$. Therefore it
deserves special attention. Surprisingly it gives 
rise to a deformation which converts on-shell new
minimal supergravity into old minimal supergravity with 
local $R$-invariance. This unusual feature is possible because
fields which are gauge fields in new minimal supergravity mutate
through the deformation to auxiliary fields or disappear for $g\neq 0$
completely from the theory after suitable local field
redefinitions. The details of the computation are in principle
straightforward but nevertheless somewhat involved. They will be given
elsewhere \cite{elsewhere}. Here I only illustrate
the underlying mechanism in a nonsupersymmetric toy model in flat
space. 

The toy model involves
two ordinary abelian gauge (vector) fields, denoted by $A_\mu$
and $a_\mu$, and a 2-form gauge potential whose components are
denoted again by $t_{\mu\nu}$. $a_\mu$ and $t_{\mu\nu}$ play roles 
analogous to the $R$-gauge field and the 2-form gauge potential 
in new minimal supergravity. The Lagrangian of the
toy model is analogous to the ``bosonic part'' of 
the sum of (\ref{act12}) and (\ref{act4}). It is given by 
\beq L=\s0 12\, H_{\mu\nu\rho}H^{\mu\nu\rho}
-\s0 23\, \ep^{\mu\nu\rho\sigma}a_\mu H_{\nu\rho\sigma}
-\s0 14\, F^{\mu\nu}F_{\mu\nu}\label{co12}\eeq
where $H_{\mu\nu\rho}=3\6_{[\mu}t_{\nu\rho]}$ and
$F_{\mu\nu}=2\6_{[\mu}A_{\nu]}$ are the field strengths
of $t_{\mu\nu}$ and $A_\mu$ respectively and all indices refer
to flat space (hence, {\em here} we work with
$\ep^{\mu\nu\rho\sigma}\in\{0,1,-1\}$).
Observe first that the field redefinition
$a'_\mu=2a_\mu+\s0 14 \ep_{\mu\nu\rho\sigma}H^{\nu\rho\sigma}$
casts (\ref{co12}) in the simpler form
\beq
L=-\ep^{\mu\nu\rho\sigma}a'_\mu \6_\nu t_{\rho\sigma}
-\s0 14\, F^{\mu\nu}F_{\mu\nu}\ .\label{co14}\eeq
The toy model is evidently
invariant under the gauge resp. BRST transformations
\beq\ba{c|c|c|c|c|c|c|c}
\Phi & t_{\mu\nu}  & A_\mu  & a'_\mu & Q_\mu  & C & c&  Q\\
\hline\rule{0em}{3ex}
s\Phi & \6_\nu Q_\mu -\6_\mu Q_\nu & \6_\mu C & \6_\mu c &
\6_\mu Q & 0 & 0 & 0
\ea
\label{co15}\eeq
The first order deformation
$\cS^{(1)}$ of the toy model analogous to the type-Vc-term
in table 8.1 is easily verified to be
\[
\cS^{(1)}=\int d^4x\, (-\s0 12\, t_{\mu\nu}F^{\mu\nu}+A^{*\mu}Q_\mu
+C^*\, Q)\]
and evidently nontrivial because $t_{\mu\nu}F^{\mu\nu}$ does
not vanish on-shell modulo a total derivative.
Elementary computations show that an $\cS^{(2)}$
satisfying (\ref{co0a}) is simply 
\[ \cS^{(2)}=-\s0 14 \int d^4x\, t_{\mu\nu} t^{\mu\nu}\]
and that
\[ \cS_g=\cS+g\, \cS^{(1)}+ g^2\cS^{(2)}\]
solves the master equation where $\cS$ is the original
solution whose integrand is $L-(s\Phi^A)\Phi^*_A$.
The deformed Lagrangian $L_g$ and BRST transformations
$s_g$ are now easily read off from $\cS_g$,
\bea & &L_g=
-\ep^{\mu\nu\rho\sigma}a'_\mu\6_\nu t_{\rho\sigma}
-\s0 14\, (F_{\mu\nu}+gt_{\mu\nu})(F^{\mu\nu}+gt^{\mu\nu}),
\label{co20}\\[3ex]
& &\ba{c|c|c|c|c|c|c|c}
\Phi & t_{\mu\nu}  & A_\mu  & a'_\mu & Q_\mu  & C & c & Q\\
\hline\rule{0em}{3ex}
s_g\Phi & \6_\nu Q_\mu -\6_\mu Q_\nu & \6_\mu C+gQ_\mu & \6_\mu c &
\6_\mu Q  & -gQ & 0 & 0
\ea
\label{co21}\eea
Evidently the deformed Lagrangian depends only on $a'_\mu$ and on
$ t'_{\mu\nu}=F_{\mu\nu}+gt_{\mu\nu}$. The latter is
$s_g$-invariant and can be algebraically
eliminated for $g\neq 0$ through its equation of motion. This turns
$L_g$ into the usual action for just one
abelian gauge field,
\beq L'_{g}=-\frac 1{g^2}\, F'_{\mu\nu} F'{}^{\mu\nu},\quad
F'_{\mu\nu}=\6_\mu a'_\nu -\6_\nu a'_\mu\ .
\label{co23}\eeq

Note that we started from an action for
three gauge fields $A_\mu,a_\mu,t_{\mu\nu}$ and, 
after the deformation and
elimination of $t'_{\mu\nu}$,
ended up with an action for only one gauge field.
What happened to the other gauge fields
and gauge symmetries? One can take the following point of view:
$t_{\mu\nu}$ mutated to an auxiliary field $t'_{\mu\nu}$ 
carrying no gauge symmetry anymore, whereas $A_\mu$ 
dropped out completely due to its deformed gauge transformation
$s_gA_\mu=Q'_\mu\equiv \6_\mu C+gQ_\mu$. We may also take the 
point of view that by deforming the model we have
gauged the global shift symmetry $A_\mu\rightarrow
A_\mu+\epsilon_\mu$ of $L$. In this perspective 
$t_{\mu\nu}$ and $Q_\mu$
are the gauge and ghost fields associated with the shift symmetry
and $t'_{\mu\nu}$ is the new field strength of $A_\mu$ 
which is covariant
(in fact invariant) with respect to the gauged shift symmetry.

This mechanism is somewhat reminiscent of a familiar
implementation of duality transformations in abelian gauge
theories, see e.g.\ \cite{duality}. In particular
the first term in the Lagrangian (\ref{co20}) is
analogous to the Lagrange multiplier term that
one introduces in that approach. In fact duality
transformations relating old and new minimal supergravity
are well-known in the literature \cite{FGKVP,bible}.
It might therefore be worthwhile to study their relations
to the deformation found here. 

\mysection{Candidate gauge anomalies}\label{anos}

The candidate gauge
anomalies are obtained from the (full)
$\4s$-cohomology at total degree 5. The latter is
summarized in table 9.1,
with $\Omega_\Delta$ and $J^\mu_\tau$ as in
(\ref{Omega}) and (\ref{covcurrents}) and writing the
candidate anomalies as $\int d^4x\, e\, L_{ano}$.
Furthermore, we use (solutions of type I and Va)
\beq C=C^IT^{(K)}_I\ ,\quad F_{\mu\nu}=\F \mu\nu{I}T^{(K)}_I
\label{CF}\eeq
for nonabelian
ghost and curvature matrices constructed by means of the
respective Lie algebra representations $\{T^{(K)}_I\}$
as in (\ref{thetas}).
\[
\ba{c|ll|l}
\mbox{Type}             &  \alpha & & L_{ano}\\
\hline\rule{0em}{3ex}
\mbox{I} & \Th_K+\dots & (m_K=3) & Tr(CF_{\mu\nu}F_{\rho\sigma})
                    \ep^{\mu\nu\rho\sigma}+\dots\\
\rule{0em}{3ex}
\mbox{II} & \4C^{i_a}\7\cP_\Delta+\dots && C^{i_a}
(\5D^2\Omega_\Delta+D^2\5\Omega_\Delta)+\dots\\
\rule{0em}{3ex}
\mbox{III} & \cH \4C^{i_a}\4C^{i_b}+\dots &&
    C^{[i_a}\A {i_b]}\mu\6_{\nu}t_{\rho\sigma}
    \ep^{\mu\nu\rho\sigma}+\dots\\
\rule{0em}{3ex}
\mbox{IVa} & N_\tau\4C^{i_a}\4C^{i_b}+\dots &&
    C^{[i_a}\A {i_b]}\mu J^\mu_\tau+\dots\\
\rule{0em}{3ex}
\mbox{IVb} & N_\tau\4Q+\dots && Q_\mu J^\mu_\tau+\dots\\
\rule{0em}{3ex}
\mbox{Va} & \4C^*_{i_a}\Th_K+\dots & (m_K=2,i_a\neq\dR)
    & -\s0 12\, Tr(CF_{\mu\nu})F^{\mu\nu}_{i_a}+\dots\\
\rule{0em}{3ex}
\mbox{Vb} & \4C^{*[i_a}\4C^{i_b}\4C^{i_c}\4C^{i_d]}+\dots& 
(i_a,\ldots,i_d\neq\dR) &
   F^{\mu\nu[i_a}\A {i_b}\nu\A {i_c}\mu C^{i_d]}+\dots\\
\rule{0em}{3ex}
\mbox{Vc} & \4C^{*[i_a}\4C^{i_b}\4C^{i_c]}\4Q^*+\dots& 
(i_a,i_b,i_c\neq\dR) &
   4C^\dR F^{\mu\nu[i_a}\A {i_b}\nu\A {i_c]}\mu +\dots\\
\rule{0em}{3ex}
\mbox{Vd} & \4C^{*[i_a} \4C^{i_b]}\4Q +\dots& (i_a,i_b\neq\dR) &
      C^{[i_a}{F_{\mu\nu}}^{i_b]}t^{\mu\nu}+\dots\\
\rule{0em}{3ex}
\mbox{Ve} & \4C^*_{i_a} \4C^\dR\4Q+\dots & (i_a\neq\dR) &
   -\s0 12\,C^\dR F^{\mu\nu}_{i_a}t_{\mu\nu}+\dots\\
\multicolumn{4}{c}{}\\
\multicolumn{4}{c}{\mbox{Table 9.1:
Candidate anomalies}}
\ea
\]
Note that the candidate anomalies
of type III, IVb, Vc, Vd, Ve occur only in new minimal
supergravity. Let me now briefly discuss and comment
these results.
\ben
\item[I.]
The solutions of type I are the supersymmetric
version of the nonabelian chiral anomalies. In section \ref{completion}
it has been shown already that they exist on general
grounds, see text after (\ref{comp5}).
Remarkably they can be chosen
so as to involve neither the antifields, nor the
gravitino, nor any of the auxiliary fields.
In this form they are given by \cite{sugra}
\bea \int Tr\left\{ Cd(AdA+\s0 12A^3)
+3i\, d^4x\, e\, (\5\xi\5\lambda\, \lambda\lambda
+\xi\lambda\, \5\lambda\5\lambda)\right. \nonumber\\
\left.
+i(\xi\sigma\,\5\lambda+\lambda\,\sigma\5\xi)
(AdA+(dA)A+\s0 32A^3)\right\}
              \label{an8}\eea
where $\xi$ and $\5\xi$ are the supersymmetry ghosts,
$C$ are ghost matrices as in (\ref{CF})
and $A$, $\lambda$, $\5\lambda$ and $\sigma$
are the matrices
\beq A=dx^\mu\A i\mu T^{(K)}_i,\
\lambda_\alpha=\lambda_\alpha^iT^{(K)}_i,\
\5\lambda_\da=\5\lambda_\da^i T^{(K)}_i,\
\sigma_{\alpha\da}=dx^\mu\sigma_{\mu\, \alpha\da}. \label{AL}\eeq
\item[II.]
The solutions of type II are analogous to
candidate anomalies in non-supersymmetric Yang--Mills
theory whose integrands are of the form ``abelian ghost
$\times$ gauge invariant function".
Their complete form has been given in \cite{sugra} (see
eq. (3.12) there). In particular they include the
supersymmetric version of abelian chiral anomalies
arising from $\Omega_\Delta=a\5\lambda^i\5\lambda_i+b\5W\5W$
where $a$ is purely imaginary. It should
be remarked however in this context that {\em in new minimal
supergravity there is no abelian chiral anomaly involving the
ghost of $R$-transformations} as the corresponding solutions
are trivial. Indeed, in new minimal
supergravity $4\4C^\dR$ can be completed
to an $\4s$-invariant total form $\4Q^*$, see (\ref{f10}).
A total form solving $\4s\alpha=0$ and corresponding to
abelian chiral anomalies involving the $R$-ghost would
therefore read $\4Q^*Tr(\cF\cF)$ with $\cF$ as in (\ref{comp4}).
However, this total form is $\4s$-exact, $\4Q^*Tr(\cF\cF)=
-\4s(\4Q^*q)$ where $q$ is the total
super-Chern--Simons form satisfying $\4s\, q=Tr(\cF\cF)$, see
section \ref{completion}.
\item[III.]
The candidate anomalies of type III 
are present only in new minimal supergravity. They
can be chosen to be antifield independent
when the auxiliary $D$-fields
are used and read then in complete form \cite{sugra} 
\bea a_{i_a i_b}\int d^4x\, e\, 
\left\{C^{i_a}(\ep^{\mu\nu\rho\sigma}\A {i_b}\mu\6_\nu t_{\rho\sigma}
+\lambda^{i_b}\sigma^\mu\5\psi_\mu
+\psi_\mu\sigma^\mu\5\lambda^{i_b}+D^{i_b})\right.\nonumber\\
\left.-\A {i_a}\mu
(\lambda^{i_b}\sigma^\mu\5\xi+\xi\sigma^\mu\5\lambda^{i_b})
-i\ep^{\mu\nu\rho\sigma}\A {i_a}\mu\A {i_b}\nu
(\xi\sigma_\rho\5\psi_\sigma+\psi_\rho\sigma_\sigma\5\xi)\right\}
\label{chiral2}\eea
where the coefficients $a_{i_a i_b}$ must be antisymmetric,
\beq a_{i_a i_b}=-a_{i_b i_a}\ , \label{antisymm}\eeq
and the use of the identifications (\ref{act10}) is understood.
Due to (\ref{antisymm}) these candidate anomalies occur only
in presence of at least one abelian gauge
symmetry in addition to $R$-symmetry.
Note that they are somewhat
similar to chiral anomalies, as the first term in the
integrand of (\ref{chiral2}) can be written completely in
terms of ghosts and connection- and curvature-forms.
I remark that the total form $\cH\4Q$ occurring in table 6.1
among the type III terms
does not give rise to a candidate anomaly because it is
trivial thanks to $\cH\4Q=\4s(\s0 12\4Q^2)$.
\item[IV.]
All type-IV-candidate anomalies involve a Noether current
and are thus present only if the classical action has at least
one nontrivial global symmetry.
The solutions of type IVa have counterparts in
Yang--Mills theory and standard gravity
\cite{noetheranos,bhprl,bbh2,bbhgrav} and occur only if there are
at least two abelian gauge symmetries. The solutions of
type IVb are present only in new minimal supergravity and
involve among others the ghosts $Q_\mu$ corresponding
to $t_{\mu\nu}$. 
\item[V.]
The candidate gauge anomalies of type V disappear as soon
as one includes matter fields transforming nontrivially
under the abelian gauge transformations,
see remark in the text after equation (\ref{zusatz}). Therefore
these solutions appear to be only of academic 
interest insofar as (true) anomalies are concerned. Recall however
that abelian gauge fields couple in supergravity always to
the gravitino and to the abelian gauginos via triple vertices
present in (\ref{act4}).
\item[VI.]
Finally I stress that in new minimal supergravity {\em any}
nontrivial Noether current $j^\mu$ gives rise to a 
previously unknown candidate gauge anomaly of the form
\beq \int d^4x\, Q_\mu j^\mu+\dots\label{newano}\eeq
This follows from tables 7.1 and 9.1 because all those solutions
in table 7.1 providing the Noether currents (type IV, Va and Vb)
occur in table 9.1 multiplied with $\4Q$.
\een

\mysection{Generalizations of the results}
\label{matter}

The investigation can of course be  extended
to the case that further (matter) multiplets are present.
It is particularly easy to include
matter multiplets which (a) have QDS-structure and
(b) transform linearly under the Yang--Mills gauge group
(and under the Lorentz group), provided one can restrict
the investigation to local functionals which depend
on the undifferentiated scalar fields of these multiplets
only via formal (possibly infinite) series'.
Indeed, under these assumptions the methods and
results of the computation extend straightforwardly
to the case that matter fields are included.

Requirement (a) is for instance satisfied off-shell for
the matter multiplets most commonly used in
supergravity theories, namely
chiral multiplets $(\varphi,\chi_\alpha,F)$ with 
supersymmetry transformations given by
\[\ba{lll} \cD_\alpha\varphi=\chi_\alpha,&
\cD_\alpha \chi_\beta=\ep_{\beta\alpha}F,&
\cD_\alpha F=-\s0 12\5M \chi_\alpha,\\
\cD_\alpha\5\varphi=0,&
\cD_\alpha\5\chi_\da=-2i\cD_{\alpha\da}\5\varphi,&
\cD_\alpha \5F=2i\cD_{\alpha\da}\5\chi^\da+B_{\alpha\da}\5\chi^\da
-4\lambda^i_\alpha\delta_i\5\varphi.
\ea\]
This implies indeed that the $D_\alpha$-representation on chiral
multiplets has QDS-structure off-shell \cite{phd,glusy}.
The result for the restricted BRST cohomology in presence
of chiral matter multiplets (under the above-mentioned assumptions)
is then easily obtained from section \ref{rest}: the
only difference is that in (\ref{cp13b})
$\cA_\Delta$ can also depend on the $\5\varphi$ whereas
$\cB_\Delta$ now also involves $\varphi$, $\chi$, $F$,
$\5\varphi$, $\5\chi$, $\5F$ and their
super-covariant derivatives, see \cite{sugra} for details.

The results for the full BRST cohomology also change only slightly,
provided one uses a standard action for the chiral
multiplets. For instance, in the simplest case the
equations of motion reduce at linearized level to
$\Box\varphi\sim 0$, $\cD_-^+\chi\sim 0$,
$F\sim 0$. It is then easy to verify that the
$D_\alpha$-representation on the chiral multiplets
has QDS-structure even on-shell. Indeed this representation 
decomposes into
singlets given by the undifferentiated $\5\varphi$'s
and the following $(D)$-multiplets:
\[
\{(D_+^+)^q\varphi,(D_+^+)^q\chi\},\
\{(D_+^+)^q\5\chi,-2i(D_+^+)^{q+1}\5\varphi\}\quad (q=0,1,\ldots)
\]
where a notation as in appendix \ref{appqds} was used.
Section \ref{full} then provides straightforwardly
the complete results for the BRST cohomology 
in presence of chiral matter fields: one must just count
$(D_+^+)^q\varphi$, $(D_+^+)^q\5\varphi$, $(D_+^+)^q\chi$ and
$(D_+^+)^q\5\chi$ among the $\7\cT$'s,
add $\5\varphi$ to the arguments
of $\cA_\Delta$ in equation (\ref{f14}), and discard
$\4C^*_{i_a}$ whenever there is a matter multiplet which
transforms nontrivially under the $i_a$th abelian gauge
symmetry.

The inclusion of linear multiplets is slightly more
involved. Namely these multiplets will give rise to some
additional cohomology as they contain 
2-form gauge potentials.
Nevertheless the results of this paper can be easily adapted
to models containing linear multiplets because one can 
show that the $D_\alpha$-representation
on such multiplets has QDS-structure too
(for the standard supersymmetry transformations
and actions). As a consequence, the presence of linear multiplets
effects the BRST cohomology in a manner very similar
to the effect that the 2-form gauge potential has
in new minimal supergravity.

The results also generalize straightforwardly
to  supergravity theories with more complicated actions (containing
for instance higher powers in
the curvatures), provided these actions can
be viewed as a continuous deformation of the simple
action with Lagrangian (\ref{f1}) in the sense
of \cite{bh} and section \ref{counter}. Namely then one
can use standard arguments from spectral sequence techniques
to relate the BRST cohomology in the more complicated
theories to that in the simple ones considered here \cite{brandt}. 
In particular the BRST cohomology 
can at most shrink but not grow
when going to a more complicated theory by a deformation.
Therefore there are for instance not more candidates for on-shell
counterterms, nontrivial deformations
or gauge anomalies than in the simple theories.
The explicit form of the solutions to the cohomological
problem may of course change.

\mysection{Note on topological aspects}\label{top}

Let me finally briefly discuss a topic which was neglected so
far and concerns `topological' solutions
to (\ref{i1}), i.e. solutions which are locally but not
globally of the form $s\eta_4+d\eta_3$. Such solutions
correspond to local total forms which are locally but not
globally $\4s$-exact and are called topological total forms
in the following. 

As mentioned already in section \ref{reduce}, the $\4s$-cohomology
is locally trivial on total forms $\alpha(\cU,\cV)$. However,
in general it is not globally trivial because the manifold of the
$\cU$'s and $\cV$'s has a nontrivial De Rham cohomology which
gives rise to topological total forms $\alpha(\cU,\cV)$.
As we are considering {\em local} total forms, this
De Rham cohomology boils down to the product of the
De Rham cohomology of $GL(4)$ carried by the
undifferentiated vierbein fields (see \cite{bbhgrav} for details)
and of the usual De Rham cohomology of the spacetime manifold.
The latter comes here into play because
the $x^\mu$ and $dx^\mu$ count among the
$\cU$'s and $\cV$'s. To give an example,
the De Rham cohomology of $GL(4)$ gives rise
to the following topological total form with total degree 3,
\[ \alpha_{top}(\cU,\cV)=Tr(\4v^3),
\quad \4v_a{}^b=\Viel a\mu\4s \viel b\mu\ .\]
The 3-form contained in
$\alpha_{top}$ is a topological conservation law
of first order in the terminology of section \ref{noether}
as it is $d$-closed but in general only locally
but not globally $d$-exact. It is given by
\[ j_{top}=Tr(v^3),
\quad v_a{}^b=\Viel a\mu d \viel b\mu\ .\]
This topological conservation law is similar to (though
different from) the one discussed
already in \cite{finkelstein}.

Using now the K\"unneth formula (\ref{kunneth}), one can construct further
topological forms involving $\alpha_{top}$ by multiplying it
with nontrivially
$\4s$-invariant total forms $\alpha(\cW)$. For instance,
in new minimal supergravity
a topological total form with total degree 4 is given by the
product $\alpha_{top}\4Q^*$ with $\4Q^*$ as 
in section \ref{step1}.
The 4-form contained in $\alpha_{top}\4Q^*$ is thus a topological
solution to (\ref{i1}) with ghost number 0 and has the form
\[ \omega_4= 4 Tr(v^3)\, A^\dR+\ldots\ ,\quad
A^\dR=dx^\mu\A \dR\mu\ . \]

Another well-known source for topological solutions
is of course the possible nontriviality of the fiber bundles
associated with the gauge fields. Prominent examples
for such solutions are the
polynomials $Tr(F^2)$ in the curvature 2-forms $F$
providing characteristic classes. 

\mysection{Conclusions}\label{conclusion}

Let me finally summarize and comment the main results of the paper.

{\em Gauge invariant actions:}
We have shown that in old minimal
supergravity the most general local action, invariant under
the standard gauge transformations, emerges from the density 
formula (\ref{act2}). This {\em proves} that in this case
the most general 
gauge invariant local action can be constructed
from superspace integrals \`a la \cite{wessbagger}.
In contrast, in new minimal supergravity there are
a few additional terms, given in (\ref{act12}) and
(\ref{act13}), which
cannot be written as superspace integrals without further ado
(one such term for each abelian gauge symmetry). 
The presence of these ``exceptional" terms
in new minimal supergravity is crucial because one
of them (corresponding to the local $R$-symmetry)
is just the supersymmetrized Einstein--Hilbert action,
whereas the others contain Fayet--Iliopoulos terms
relevant for spontaneous supersymmetry
breaking \cite{FI}. As all these exceptional contributions contain
Chern--Simons like terms involving the 2-form gauge potential
and abelian gauge fields, the situation
is somewhat similar to standard (non-supersymmetric)
gravity. There the Chern--Simons contributions are the only
exceptions to the rule that invariant actions are
integrated scalar densities of the form ``vielbein determinant
$\times$ scalar function of the tensor fields" \cite{grav}.

{\em Consistent deformations:}
The gauge transformations
are extremely stable under continuous deformations of
the type described in the introduction. 
In old minimal supergravity the gauge transformations cannot be
consistently deformed in a continuous and nontrivial manner 
whenever the Yang--Mills gauge group is semisimple%
\footnote{I stress again that in this form the statement
applies to the formulation of the theory {\em with} auxiliary fields.
The elimination of the auxiliary fields can of course modify
the on-shell gauge transformations. For instance, adding the
cosmological contribution (\ref{act5}) to the standard supergravity 
Lagrangian (\ref{act3}) (without local $R$-symmetry, i.e.\ for
$\lambda^\dR\equiv D^\dR\equiv 0$) provides
a consistent deformation of old minimal supergravity 
with deformation parameter $m$ which does not change the gauge
transformations as long as one keeps the auxiliary fields. However,
upon elimination of the auxiliary field $M$,
this deformation modifies the 
supersymmetry transformation of $\psi_\mu$ on-shell by 
a term proportional to $i\5m \5\xi\5\sigma_\mu$, due to the
presence of $M$ in the off-shell supersymmetry 
transformation of $\psi_\mu$. This modification
indicates of course spontaneous supersymmetry breaking
related to the cosmological constant and gravitino mass term
introduced by (\ref{act5}).}.
As in ordinary (non-supersymmetric) gravity \cite{bbhgrav},
there may be nontrivial deformations in presence of abelian 
gauge symmetries but it seems that all of them are well-known,
gauging either global symmetries in the standard way or deforming
abelian to standard nonabelian gauge multiplets.
In new minimal supergravity the
situation is slightly different. In particular there
exists an unusual deformation which converts new 
into old minimal supergravity with gauged $R$-transformations
and is reminiscent of a duality transformation, see
\cite{elsewhere} for details.

{\em On-shell counterterms:}
We have obtained the complete
lists of the possible counterterms that are gauge invariant on-shell 
for old and new minimal supergravity
(these lists are actually overcomplete as they 
still contain terms which reduce on-shell to surface integrals). 
It turns out that
apart from very few exceptions all these on-shell counterterms 
can actually be completed to off-shell invariants
by means of the auxiliary fields. In old minimal supergravity
exceptions occur only if there are (i) at least one 
nontrivial Noether current {\em and} one abelian gauge symmetry, or 
(ii) at least three abelian gauge symmetries. A similar
result holds in new minimal supergravity.

{\em Candidate gauge anomalies:}
We have classified the possible
gauge anomalies completely (up to ``topological anomalies"). The result
indicates in particular that supersymmetry itself is not anomalous
in minimal supergravity
because all the candidate anomalies have counterparts
in the corresponding non-supersymmetric theories. 
In other words, supersymmetry
does not introduce new types of candidate anomalies. 
In old minimal supergravity
the supersymmetrized version (\ref{an8})  of the familiar
nonabelian chiral anomalies 
exhausts the candidate gauge anomalies
whenever the Yang--Mills gauge group is semisimple.
All other candidate gauge anomalies require thus
the presence of at least one abelian gauge symmetry, as in standard
Einstein--Yang--Mills theories \cite{grav,bbhgrav}.
In new minimal supergravity there are special candidate anomalies
due to the presence of the two-form gauge potential. 
They are given by (\ref{chiral2}) and (\ref{newano}). The
former are somewhat similar to chiral anomalies and had been found
already in \cite{sugra}. The
latter were previously unknown. They
correspond one-to-one to the nontrivial Noether currents
(if any) and involve the ghosts associated with the 2-form gauge
potential.

{\em Dynamical conservation laws:}
New minimal supergravity possesses
precisely one dynamical conservation law of order 3,
i.e.\ one dynamically conserved 1-form (up to trivial ones).
Old minimal supergravity does not admit such a conservation law
because the gauge transformations are irreducible in this case. The
dynamical conservation laws of order 2
(i.e. the nontrivial conserved 2-forms) are exhausted by supersymmetric
completions of the duals of abelian curvature forms, both in
old and new minimal supergravity. 
The results for the dynamical conservation laws of first order, 
i.e.\ for the Noether currents and the corresponding global symmetries
are incomplete. Nevertheless we found both well-known
global symmetries (global $R$-invariance of old minimal supergravity,
invariance under global $SO(N)$-rotations of $N$ abelian gauge 
multiplets), as well as further (possibly previously 
unknown) global symmetries 
of new minimal supergravity when it is coupled to abelian gauge multiplets.
The latter global symmetries transform for instance 
the 2-form gauge potential among
others into the dual of an abelian field strength.
It is likely, though not proved,
that these are all the inequivalent and nontrivial global
symmetries of minimal supergravity in absence of matter multiplets.

Finally I stress again that these results hold thanks to the
``QDS structure" \cite{phd,glusy} of old and new minimal supergravity.
This structure refers to the representation of the
linearized supersymmetry algebra, or rather its
subalgebra (\ref{qds2}) on tensor fields and holds both off-shell 
and on-shell. The QDS structure remains intact even
when chiral matter multiplets are included, at least for the
standard actions. In this case the results remain therefore 
essentially unchanged, see section \ref{matter} for details.
The inclusion of linear multiplets is also straightforward even
though these multiplets will give rise to some additional cohomology, 
similar to the effect that the presence of the 2-form gauge 
potential has in new minimal supergravity.

It can be shown \cite{phd} that the QDS structure of minimal supergravity
is somewhat related to the so-called constraints satisfied
by the torsions and curvatures occurring in the covariant supergravity
algebra (\ref{s2}). Other constraints, such
as those realized in non-minimal supergravity \cite{nonmin}, could
destroy the QDS structure and therefore might
lead to different cohomological 
results. In particular candidate anomalies for supersymmetry itself
may be present in such a case. The arguments in \cite{gates} 
suggest that such candidate anomalies might occur 
in non-minimal supergravity, but this
has not yet been confirmed by cohomological means
(see however \cite{sugra} for some results on the 
restricted BRST cohomology in non-minimal supergravity 
supporting this conjecture).

\section*{Acknowledgements}

I am grateful to Norbert Dragon for many
enlightening discussions and pieces of
advice, in particular during the work on \cite{phd}.
I am also indebted to Marc Henneaux, James Gates, 
Jordi Par\'{\i}s, Kostas Skenderis and Antoine Van Proeyen
for useful discussions and remarks. This work was supported in part by
grants of the research council of the K.U. Leuven (DOC) and of
the Spanish ministry of education and science (MEC), and by 
the TMR programme ERBFMRX--CT96--0045.

\appendix
\mysection{Conventions}\label{conventions}

The conventions concerning the BRST algebra
(BRST operator, antibracket etc.)
agree with those used in \cite{ten}.

\subsection{Lorentz algebra}\label{app1}

Minkowski metric, $\ep$-tensors:
\beann & &\eta_{ab}=diag(1,-1,-1,-1),\qd
\ep^{abcd}=\ep^{[abcd]},\qd \ep^{0123}=1,\\
& &\ep^{\alpha\beta}=-\ep^{\beta\alpha},\qd
\ep^{\da\dbe}=-\ep^{\dbe\da},\qd
\ep^{12}=\ep^{\dot 1\dot 2}=1,\\
& &\ep_{\alpha\gamma}\ep^{\gamma\beta}=\delta_\alpha^\beta=diag(1,1),\qd
\ep_{\da\dg}\ep^{\dg\dbe}=\delta_\da^\dbe=diag(1,1)            \eeann
$\sigma$-matrices: $\sigma^a_{\alpha\da}$
($\alpha$: row index, $\da$: column index):
\[
\sigma^0=\left( \begin{array}{rr}1&0\\ 0&1\end{array}\right),\qd
\sigma^1=\left( \begin{array}{rr}0&1\\ 1&0\end{array}\right),\qd
\sigma^2=\left( \begin{array}{rr}0&-i\\ i&0\end{array}\right),\qd
\sigma^3=\left( \begin{array}{rr}1&0\\ 0&-1\end{array}\right)  \]
$\5\sigma$-matrices:
\beann \5\sigma^{a\, \da\alpha}=
\ep^{\da\dbe}\ep^{\alpha\beta}\sigma^a_{\beta\dbe}     \eeann
$\sigma^{ab},\5\sigma^{ab}$-matrices:
\beann \sigma^{ab}=
\s0 14(\sigma^a\5\sigma^b-\sigma^b\5\sigma^a),\qd
\5\sigma^{ab}
=\s0 14(\5\sigma^a\sigma^b-\5\sigma^b\sigma^a) \eeann
Lorentz ($SL(2,C)$) transformations:
\beq
l_{ab}V_c=-2\eta_{c[a}V_{b]},\
l_{\alpha\beta}\psi_\gamma=-\ep_{\gamma(\alpha}\psi_{\beta)},\
l_{ab}=\sigma_{ab}{}^{\alpha\beta}l_{\alpha\beta}
-\5\sigma_{ab}{}^{\da\dbe}\5l_{\da\dbe}.
\label{ltrafo}\eeq

\subsection{Spinors, Grassmann parity and complex conjugation}
\label{app2}

We work with two-component Weyl spinors. Undotted and dotted
spinor indices $\alpha,\da$ distinguish
the $(\s0 12,0)$ and $(0,\s0 12)$ representations
of $SL(2,C)$, related by complex conjugation.

Raising and lowering of spinor indices:
\beann \psi_\alpha=\ep_{\alpha\beta}\psi^\beta,\qd
\psi^\alpha=\ep^{\alpha\beta}\psi_\beta,\qd
\5\psi_\da=\ep_{\da\dbe}\5\psi^\dbe,\qd
\5\psi^\da=\ep^{\da\dbe}\5\psi_\dbe\ .
                                                             \eeann

Contraction of spinor indices:
\beann \psi\chi:=\psi^\alpha\chi_\alpha,\qd
\5\psi\5\chi:=\5\psi_\da\5\chi^\da,\qd
\psi^\ua\chi_\ua:=\psi\chi+\5\psi\5\chi.
\eeann

Lorentz vector indices in spinor notation:
\beann V_{\alpha\da}=\sigma^a_{\alpha\da} V_a\ ,\qd
V^{\da\alpha}=\5\sigma_a^{\da\alpha} V^a. \eeann

The Grassmann parity $\ep(X)$ of a variable
(field, antifield, differential or spacetime coordinate)
or an operator 
is determined by the number of its spinor indices, its ghost
number ($\gh$) and its form degree ($\deg$) according to
\[ \ep(X_{\alpha_1\ldots\alpha_n}^{\da_1\ldots\da_m} )=
m+n+\gh(X)+\deg(X)\qd (mod\ 2).                        \]
The Grassmann parity of the variables, denoted
collectively by $Z^i$, determines their statistics,
\beann Z^iZ^j=(-)^{\ep(Z^i)\, \ep(Z^j)}Z^jZ^i.  \eeann
Complex conjugation of a variable or an operator $X$
is denoted by $\5X$.
Complex conjugation of products of variables and operators is defined by
\beann \overline{XY}=(-)^{\ep( X)\, \ep( Y)}\5X\ \5Y. \eeann
In particular this implies
\beann\overline{\6/\6Z}=(-)^{\ep(Z)}\6/\6\5Z\ .
\eeann

\mysection{Gauge covariant algebra}
\label{appA}

This appendix spells out explicitly the realization of
the gauge covariant algebra (\ref{s2}) in old
and new minimal (Poincar\'e) supergravity. The super-covariant 
derivative $\cD_a$ is defined according to
\beq \cD_a=\Viel a\mu(\6_\mu-\A I\mu\delta_I
-\RS \alpha\mu\cD_\alpha-\5\psi_{\mu\da}\5\cD^\da)\label{ca-1}\eeq
where
\[\{\A I\mu\}=\{\A i\mu\, ,\ \spin {ab}\mu\, :\, a>b\}\]
contains the Yang--Mills gauge fields $\A i\mu$ and the
spin connection
\bea
\spin {ab}\mu&=&E^{a\nu}E^{b\rho}(\omega_{[\mu\nu]\rho}
-\omega_{[\nu\rho]\mu}+\omega_{[\rho\mu]\nu}),
\nonumber\\
\omega_{[\mu\nu]\rho}&=&e_{\rho a}\6_{[\mu}\viel a{\nu]}
-i\psi_{\mu}\sigma_\rho\5\psi_{\nu}
+i\psi_{\nu}\sigma_\rho\5\psi_{\mu}\ .
\label{ca2}\eea

The nonvanishing $\Gg IAB$ occurring in  (\ref{s2})
are read off from
\beq\ba{lll} [l_{ab},\cD_c]=2\eta_{c[b}\cD_{a]},&
[l_{ab},\cD_\alpha]=-{\sigma_{ab\,\alpha}}^\beta\cD_\beta,&
[l_{ab},\5\cD_\da]=\5\sigma_{ab}{}^\dbe{}_\da\5\cD_\dbe,\\
{}[\delta_\dR,\cD_a]=0,&
[\delta_\dR,\cD_\alpha]=-i\,\cD_\alpha,&
[\delta_\dR,\5\cD_\da]=i\, \5\cD_\da .
\ea\label{ca0}\eeq
We use a formulation
of old minimal supergravity with torsions $\T ABC$
and curvatures $\F ABI$ as in table B.1 where
$\T \mu\nu\alpha$ and $\F \mu\nu{I}$ are given in terms of the
gauge fields and the remaining torsions and curvatures according
to
\bea
\F \mu\nu{I}&=&
2(\6_{[\mu}\A I{\nu]}+\s0 12\f JKI\A J\mu\A K\nu
            +\viel c{[\mu}\RS \ua{\nu]}\F \ua{c}I
            +\s0 12\RS \ua\mu\RS \ube\nu\F \ua\ube{I}),
                                                        \label{ca3}\\
\T \mu\nu\alpha&=&
2(\6_{[\mu}\RS \alpha{\nu]}+\viel c{[\mu}\RS \ube{\nu]}\T \ube{c}\alpha
                        +\RS \beta{[\mu}\A I{\nu]}\Gg I\beta\alpha).
                                                      \label{ca4}\eea
\[
\ba{c|c|c|c|c}
AB & ab & \da b & \da\dbe & \alpha\dbe \\
\hline\rule{0em}{3ex}
\T ABc & 0 & 0 & 0 & 2i\sigma^c_{\alpha\dbe} \\
\rule{0em}{3ex}
\T AB\gamma & \Viel a\mu\Viel b\nu\T \mu\nu\gamma &
           \s0 i8 M\ep^{\gamma\alpha}\sigma_{b\,\alpha\da} & 0 & 0 \\
\rule{0em}{3ex}
\T AB\dg & -\Viel a\mu\Viel b\nu\5T_{\mu\nu}{}^\dg &
              i\, (\delta_\da^\dg B_b
               +B^c\5\sigma_{cb}{}^\dg{}_\da) & 0 & 0 \\
\rule{0em}{3ex}
\F ABi & \Viel a\mu\Viel b\nu\F \mu\nu{i} &
       i\lambda^{i\,\alpha}\sigma_{b\,\alpha\da}  & 0 & 0 \\
\rule{0em}{3ex}
\F AB{cd} & \Viel a\mu\Viel b\nu\F \mu\nu{cd} &
       iT^{cd\alpha}\sigma_{b\, \alpha\da}
       -2i\sigma^{[c}_{\alpha\da} T^{d]}{}_b{}^\alpha &
       -M\5\sigma^{cd}{}_{\da\dbe} &
       2i \ep^{abcd}\sigma_{a\alpha\dbe}B_b \\
\multicolumn{5}{c}{}\\
\multicolumn{5}{c}{\mbox{Table B.1:
Torsions and curvatures in old minimal supergravity}}
\ea
\]

The explicit realization of
$\cD_\alpha$ and $\5\cD_\da$ on the tensor fields
(\ref{s3}) can be obtained \`a la \cite{covsugra} from an analysis
(``solution") of the Bianchi identities
\bea
0&=&\csum {ABC}{35}
(\cD_A\T BCD+\T ABE\T ECD+\F ABI\Gg ICD),
\label{ca5}\\
0&=&\csum {ABC}{35}
(\cD_A\F BCI+\T ABD\F DCI)
\label{ca6}\eea
where
\[ \csum {ABC}{35}X_{ABC}=(-)^{\ep_A\ep_C}X_{ABC}
+(-)^{\ep_B\ep_A}X_{BCA}+(-)^{\ep_C\ep_B}X_{CAB}\ .\]
Using the notation of (\ref{bas1}) and (\ref{bas2})
one finds in particular
\bea
\cD_\alpha M &=& \s0 {16}3\,(S_\alpha-i\lambda_\alpha^\dR),
\\                                           
\cD_\alpha \5M & =& 0,\label{ca8}\\
\cD_\alpha B_{\beta\dbe}
&=& \s0 13\ep_{\beta\alpha}(\5S_\dbe+4i\5\lambda_\dbe^\dR)
                     -\5U_{\alpha\beta\dbe},  \label{ca9} \\
\cD_\alpha \lambda_\beta^i &=& i\ep_{\alpha\beta}D^i
                             +\G \alpha\beta{i},\\
\cD_\alpha\5\lambda_\da^i & =& 0,  \label{ca11}\\
\cD_\alpha D^i &=&
\cD_{\alpha\da}\5\lambda^{i\da}+
\s0 {3}2iB_{\alpha\da}\5\lambda^{i\da}\ .\label{ca12}\eea
The $\cD_\alpha$-transformations of $\T ab\beta$,
$\T ab\dbe$ and $\F abI$
are easily obtained from the identities
(\ref{ca5}) and (\ref{ca6}) with indices $(ABC)\equiv(\alpha ab)$
and are therefore not spelled out here (the linearized version of these
transformations is given in appendix \ref{appqds}).

The realization of the algebra (\ref{s2}) in new minimal supergravity
can be obtained from the above formulae using
the identifications
\beq M\equiv 0, \qd B^a\equiv\s0 16\ep^{abcd}H_{bcd},\qd
D^\dR\equiv-\s0 14\, (\cR+H_{abc}H^{abc}),\qd
\lambda_\alpha^\dR\equiv -iS_\alpha\label{ca13}\eeq
with $H_{\mu\nu\rho}$ as in (\ref{identify}).
$H_{abc}$ satisfies
\beq \ep^{abcd}\,\cD_{a}H_{bcd}=0.\label{ca15}\eeq
This can be verified using (\ref{ca-1}) and
\beq \cD_\alpha\4H_{\beta\dbe}=
\ep_{\alpha\beta}\5S_\dbe-\5U_{\alpha\beta\dbe}\ ,
\quad \4H^a=\s0 16\ep^{abcd}H_{bcd}
\label{ca16}\eeq
which is consistent with (\ref{ca9}) and (\ref{ca13}).

\mysection{QDS-structure of minimal supergravity}\label{appqds}
\subsection{Definition of QDS-structure}

It will now be shown, both  for
old and for new minimal supergravity, that the off-shell and the
on-shell representations of the
linearized supersymmetry algebra (\ref{c10}) on tensor fields
have `QDS-stucture' in the terminology of \cite{glusy}.
Let me first repeat the definition
of this structure.
It refers to the representation
of the subalgebra
\beq \{D_\alpha ,D_\beta\}=0
\label{qds2}\eeq
of (\ref{c10}) on the independent tensor fields in the theory.
To analyze it, we use the same notation $\cT_n^m$ as in section
\ref{basis}
for a Lorentz-irreducible multiplet of tensor fields whose
components carry $n$ undotted and $m$ dotted spinor indices
and are completely symmetric in them respectively.
We now define operations $D_+$ and $D_-$
\beq D_+\cT^m_n\equiv
\{D_{(\alpha_0}\cT_{\alpha_1\cdots\alpha_n)}^{\da_1\cdots\da_m}\},
\quad
D_-\cT^m_n\equiv \{n
D^{\alpha_n}\cT_{\alpha_1\cdots\alpha_n}^{\da_1\cdots\da_m}\}.
\label{qds1}\eeq
The possible indecomposable representations of (\ref{qds2})
(``$D_\alpha$-multiplets")
have been determined in \cite{glusy}: there are
singlets $(S)\equiv S^m_n$,
`quartet-representations' $(Q)\equiv\{Q^{(0)}{\,}_n^m,
Q^{(-)}{\,}_{n-1}^m,
Q^{(+)}{\,}_{n+1}^m,Q^{(+-)}{\,}_n^m\}$ which degenerate to triplets
in the case $n=0$, and `zig-zag representations'
$(Z)\equiv\{Z^{(0)}{\,}^m_n,\ldots,Z^{(k)}{\,}^m_{n+k}\}$ with an
arbitrary number $k$ of components. It should be remarked
that the linearized supersymmetry algebra
(\ref{c10}) itself does not rule out any of
these representations. The representation of
(\ref{qds2}) is said to have QDS-structure if it decomposes
completely into singlets $S^m_0$ (i.e. singlets
without undotted spinor index), $(Q)$-multiplets (of any kind) and
very special $(Z)$-multiplets, called $(D)$-multiplets,
of the form $\{D^{(0)}{\,}_n^m, D^{(+)}{\,}_{n+1}^m\}$.
The properties of these multiplets are summarized in table C.1.
I note that one has $Q^{(+-)}=\s0 12\, D^2\, Q^{(0)}$.
\[
\ba{c|c|c|c}
\mbox{Multiplet} & \cT & D_-\cT & D_+\cT\\
\hline\rule{0em}{3ex} 
(Q) & Q^{(0)}{\,}_n^m & Q^{(-)}{\,}_{n-1}^m & Q^{(+)}{\,}_{n+1}^m\\
\rule{0em}{3ex}
    & Q^{(-)}{\,}_{n-1}^m & 0 & -n\, Q^{(+-)}{\,}_n^m\\
\rule{0em}{3ex}
    & Q^{(+)}{\,}_{n+1}^m & (n+2)\, Q^{(+-)}{\,}_n^m & 0\\
\rule{0em}{3ex}
    & Q^{(+-)}{\,}_n^m & 0 & 0\\
\rule{0em}{3ex}
(D) & D^{(0)}{\,}_n^m & 0 & D^{(+)}{\,}_{n+1}^m\\
\rule{0em}{3ex}
    & D^{(+)}{\,}_{n+1}^m & 0 & 0\\
\rule{0em}{3ex}
(S) & S^m_0 & 0 & 0\\
\multicolumn{4}{c}{}\\
\multicolumn{4}{c}{\mbox{Table C.1:
$D_\alpha$-multiplets in QDS-theories}}
\ea
\]

{\em Remark:} Note that the $D_\alpha$-multiplets defined
above are actually multiplets of multiplets because their
components are complete Lorentz (resp.\ $SL(2,C)$)
multiplets.

\subsection{Off-shell $D_\alpha$-multiplets}\label{offqds}

For old minimal supergravity,
the $D_\alpha$-transformations of all 
those tensor fields which do not
carry super-covariant derivatives are listed
in table C.2. The $D_\alpha$-transformations
of their super-covariant derivatives follow then
from the algebra (\ref{c10}), using the
linearized Bianchi identities (\ref{bas8})--(\ref{bas12}).
The $D_\alpha$-transformations
in new minimal supergravity are obtained from table
C.2 using the identifications (\ref{ca13}) and the
additional Bianchi identity (\ref{ca15}).
\[
\ba{c|c|c||c|c|c}
\cT & D_-\cT & D_+\cT & \cT & D_-\cT & D_+\cT\\
\hline\rule{0em}{3ex}
B  & -\s0 23(\5S+4i\5\lambda^\dR) & -\5U & & & \\
\rule{0em}{3ex}
M  & 0 &  \s0 {16}3(S-i\lambda^\dR) & \5M & 0 & 0\\
\rule{0em}{3ex}
S  & \s0 12(\cR+3iD_-^-B) & iG^\dR&\5S & 0 & -\s0 38iD^+_+\5M \\
\rule{0em}{3ex}
W  & 6iD_+^-B-4iG^\dR & -\s0 12X & \5W  & 0 & 0 \\
\rule{0em}{3ex}
U  & 2i(D_-^+B+\5G^\dR) & iD_+^+B-\s0 12Y &
\5U  & \s0 34iD_+^+\5M & 0 \\
\rule{0em}{3ex}
\cR  & 0 & 4iD^-_+\5S & & & \\
\rule{0em}{3ex}
Y  & 10iD_-^+\5U+6iD^+_+\5S & -2iD^+_+\5U & & & \\
\rule{0em}{3ex}
X  & -20iD^-_+\5U & 0 &\5X  & 0 &-8iD^+_+\5W \\
\rule{0em}{3ex}
\lambda^i  & 2iD^i & G^i & \5\lambda^i  & 0 & 0 \\
\rule{0em}{3ex}
G^i  & -6iD_+^-\5\lambda^i & 0 &
\5G^i  & 0 & 2iD^+_+\5\lambda^i \\
\rule{0em}{3ex}
D^i  & 0 & D_+^-\5\lambda^i & & & \\
\multicolumn{6}{c}{}\\
\multicolumn{6}{c}{\mbox{Table C.2:
Off-shell $D_\alpha$-transformations in old minimal supergravity}}
\ea
\]
It is now straightforward (though somewhat tedious)
to verify the QDS-structure
of old and new minimal supergravity. Indeed one finds that all the
tensor fields are either $D_\alpha$-singlets $S^m_0$
or group into $(D)$- and $(Q)$-multiplets.
More precisely, the singlets are exhausted by
\[ (S):\quad \5M,\, \5W,\, \5\lambda^i\]
and the $(D)$-multiplets are, using the notation
$\{D^{(0)},\,  D^{(+)}\}$,
\[ (D):\quad\{(D^+_+)^q\, \5X,\, -8i(D^+_+)^{q+1}\5W\},\
\{(D^+_+)^q\, \5G^i,\, 2i(D^+_+)^{q+1}\5\lambda^i\}
\quad(q=0,1,\ldots).\]
All other multiplets are $(Q)$-multiplets.
Here I only list
their lowest components $Q^{(0)}$ (the full multiplets
are spelled out in \cite{phd}):
\[\ba{llll}
(D_+^+)^qB,& \Box^p(D_+^+)^qD_-^-B,&
\Box^p(D_+^+)^q M,& (D_+^+)^qW,\\
\Box^p(D_+^+)^q U,&
\Box^p(D_+^+)^qD_-^+U,&
\Box^p(D_+^+)^q D_+^-U,&
\Box^p(D_+^+)^q(D_+^-)^2U,\\
\Box^p(D_+^+)^q \lambda^i,&
\Box^p(D_+^+)^qD_-^+\lambda^i.& & \ea\]
The higher components of the $(Q)$-multiplets are
in general (multiplets of)
linear combinations of the tensor fields $\cT^\TIX$.
These linear combinations
are linearly independent and form together with the
singlets and the components of the $(D)$-multiplets
a basis for the tensor fields in the sense
of section \ref{basis}, equivalent to $\{\cT^\TIX\}$.
This was shown explicitly in \cite{phd}
and implies the QDS-structure of the off-shell
representation of (\ref{qds2}) in old and new minimal
supergravity.

\subsection{On-shell $D_\alpha$-multiplets}\label{onqds}

Using table C.2 and the linearized equations of motion it is
easy to verify that the $D_\alpha$-transformations of
the tensor fields (\ref{f3}) reduce on-shell to those
given in table C.3.
\[
\ba{c|c|c||c|c|c}
\7\cT & D_-\7\cT & D_+\7\cT & \7\cT & D_-\7\cT & D_+\7\cT \\
\hline\rule{0em}{3ex}
W_q & 0 & -\s0 12 X_q & \5W_q & 0 & 0\\
\rule{0em}{3ex}
\5X_q & 0 & -8i\5W_{q+1} & X_q & 0 & 0 \\
\rule{0em}{3ex}
\lambda^i_q & 0 & G^i_q & \5\lambda^i_q &0 &0\\
\rule{0em}{3ex}
\5G^i_q & 0 & 2i\5\lambda^i_{q+1} & G^i_q & 0 & 0\\
\multicolumn{6}{c}{}\\
\multicolumn{6}{c}{\mbox{Table C.3:
On-shell $D_\alpha$-transformations}}
\ea
\]
The QDS-structure of the on-shell representation of (\ref{qds2})
is evident from table C.3: the on-shell
$D_\alpha$-multiplets are singlets without
undotted spinor index and $(D)$-multiplets (there are
no $(Q)$-multiplets on-shell in this case!):
\beann (S):& & \5W_0,\quad
\5\lambda^i_0\nonumber\\
(D):& &\{W_q,-\s0 12X_q\},\ \{\5X_q,-8i\5W_{q+1}\},\
\{\lambda^i_q,G^i_q\},\ \{\5G^i_q,2i\5\lambda^i_{q+1}\}
\quad (q=0,1,\ldots )
\eeann

\mysection{Super-covariant Poincar\'e lemma}\label{CPL}

In this appendix it is proved that any $\4s$-exact
and $\cG$-invariant local total form
$f(\4\xi,\cT)$ is of the form $\4s h(\4\xi,\cT)$
except for
the abelian total curvature forms $\cF^{i_a}$,
the total curvature form $\cH$ corresponding to
$t_{\mu\nu}$ and
the ``total super-Chern 4-forms",
\bea & &f(\4\xi,\cT)=\4s \beta,
\quad \delta_If(\4\xi,\cT)=0\nonumber\\
& &\then\quad
f(\4\xi,\cT)=\4s h(\4\xi,\cT)+a_{i_a}\cF^{i_a}
+c\,\cH+d_{IJ}\cF^I\cF^J\label{lem71}\eea
where the $a_{i_a}$ and $c$ are arbitrary
constants and the $d_{IJ}$ are constant $\cG$-invariant
symmetric tensors. Furthermore, no nonvanishing
linear combination of the $\cF^{i_a}$, $\cH$ and
$d_{IJ}\cF^I\cF^J$ is $\4s$-exact in the space of
local total forms $f(\4\xi,\cT)$,
\beq a_{i_a}\cF^{i_a}
+c\,\cH+d_{IJ}\cF^I\cF^J=\4s h(\4\xi,\cT)
\quad\LRA\quad
a_{i_a}=c=d_{IJ}=0.
\label{lem72}\eeq

To prove (\ref{lem71}) I first note that
thanks to (\ref{r33}) we can assume
without loss generality that
$\beta$ does not involve
the $\cU$'s and $\cV$'s. Furthermore we can of course assume
that $f$ has a definite total degree $G$ and thus consider
\beq f(\4\xi,\cT)=\4s \beta(\cW),\quad \delta_If=0,\quad
\tot(f)=G.\label{m-1}\eeq
For $G>4$, the assertion follows immediately from
(\ref{c3}). Indeed, (\ref{m-1}) implies $\4s f=0$ and
(\ref{c3}) therefore ensures
$f=\4s h(\4\xi,\cT)$ for some local $h$ in the cases $G>4$.
Note that we can use (\ref{c3}) in the cases $G>4$
because these results are derived independently of the
above assertions, in contrast to the results for $G<4$.

We are therefore left with the cases $G\leq 4$.
Since $\beta$ has total degree $(G-1)$, it vanishes
for $G=0$ (as it does not involve
antifields and thus cannot have negative ghost number)
and is necessarily of the form $h(\cT)$ in the
case $G=1$. We conclude
\bea G=0:& & f=0,\label{m0a}\\
G=1:& & f(\4\xi,\cT)=\4s h(\cT)
\label{m0}\eea
which proves (\ref{lem71}) for $G=0,1$.

The cases $G=2,3,4$ are more involved. Using
(\ref{c2}) they can be treated by adapting a method developed in
appendix E of \cite{bbhgrav} to solve a similar problem
in ordinary gravity. To that end we define $\delta_I$ on the
$\4C^J$ and $\4Q$ according to
\beq \delta_I \4C^J=-\f IKJ \4C^K,\qd
\delta_I\4Q=0\label{m1}\eeq
i.e.\ $\4C^J$ transforms under $\delta_I$ according to the adjoint
representation of $\cG$, whereas $\4Q$ is $\cG$-invariant.
It is now easy to check that on
local total forms $\alpha(\cW)$ one has
\beq \delta_I=\{\4s,\6_I\}\qd\then\qd [\delta_I,\4s]=0
\label{m2}\eeq
where $\6_I$ is the derivative w.r.t. $\4C^I$,
\beq \6_I=\frac{\6}{\6\4C^I}\ .\label{m3}\eeq
(\ref{m2}) implies that without loss of generality we can assume
\beq\delta_I\beta=0\label{m4}\eeq
because $\delta_I f=0$ implies
$\4s(\delta_I\beta)=0$, i.e.\ any $\cG$-{\em noninvariant}
contribution to $\beta$ would have to be $\4s$-invariant and 
would thus not contribute to $f$ in (\ref{m-1}). 
Applying now $\6_I$ to (\ref{m-1}) we get,
thanks to (\ref{m2}),
\beq \4s(\6_I\beta)=0\label{m5}\eeq
since $f$ does not depend on the  $\4C^I$.
Hence $\6_I\beta$ is $\4s$-closed, has
total degree $(G-2)$ and is thus $\4s$-exact for $G=3,4$ and
constant for $G=2$ by (\ref{c2}),
\bea G=2:& & \6_I\beta=a_I=constant,\label{m6}\\
G=3:& & \6_I\beta=\4s h_I(\cT),\label{m7}\\
G=4:& & \6_I\beta=\4s \beta_I(\cW)\ ,\label{m7a}
\eea
where in (\ref{m7}) we used that $h_I$ has vanishing total degree
and thus depends only on the $\cT$ (we also used (\ref{r33}) again).

Let us first consider the case $G=2$. Since in this case
$\beta$ has total degree 1,
(\ref{m6}) implies evidently $\beta=h(\4\xi,\cT)+a_I\4C^I$.
(\ref{m4}) then requires 
$a_I=0$ unless $I$ refers to an
abelian element of $\cG$. We conclude
\bea G=2:& &\beta=h(\4\xi,\cT)+a_{i_a}\4C^{i_a}
\nn\\
&\then  & f=\4s h(\4\xi,\cT)+a_{i_a}\cF^{i_a}\ .
\label{m8}\eea

Next we turn to the case $G=3$.
In (\ref{m7}) we can assume with no loss of generality
that the $h_I$ transform under $\cG$ according to the co-adjoint
representation because this holds for $\6_I\beta$ too (as a
consequence of (\ref{m4}), due to $[\delta_I,\6_J]=\f IJK\6_K$) 
and because $\4s$ leaves the
representation invariant due to (\ref{m2}).
Applying $\6_J$ to (\ref{m7}) we thus conclude, using
(\ref{m2}) again,
\beq G=3:\quad
\6_J\6_I\beta=\f JIK h_K\ .\label{m9}\eeq
In the case $G=3$, $\beta$ has total degree
2 and is thus at most quadratic in the $\4C^I$ and
linear in $\4Q$. Hence, it is of the form
\beq G=3:\quad \beta=\7h+\4Q\7g+\4C^I\7h_I+
\s0 12 \4C^I\4C^J\7h_{JI}\label{m9a}\eeq
where $\7g$ and the $\7h$'s depend only on the $\4\xi$'s and
$\cT$'s and can be assumed to transform under $\cG$
according to their indices due to (\ref{m4}).
It is now easy to verify that
(\ref{m7}) and (\ref{m9}) imply
\bea G=3:&&
\7h_{JI}=\f JIK h_K,\qd \7h_I=\4s_{susy} h_I
\nonumber\\
&\then  & \beta=\7h+\4Q\7g-\4s(\4C^I h_I)+\cF^I h_I
\nn\\
&\then  & f=\4s(\7h+\cF^I h_I)+\cH\7g+\4Q(\4s\7g).
\label{m10}\eea
Now, since $f$ does not depend on $\4Q$, differentiation
of (\ref{m10}) w.r.t.\ $\4Q$ yields $\4s \7g=0$ which
implies $\7g=c=constant$ by (\ref{c2}) as $\7g$ has
vanishing total degree. This yields
\beq G=3:\quad f=\4s h(\4\xi,\cT)+c\,\cH,\label{m10a}\eeq
with $h=\7h+\cF^I h_I$, and
proves (\ref{lem71}) in the case $G=3$.

The case $G=4$ can be treated similarly. Applying $\6_J$ to
(\ref{m7a}) yields
\bea G=4:& & \6_J\6_I\beta=\f JIK \beta_K-\4s \, \6_J\beta_I
\label{m11}\\
&\then  &\4s\, \6_{(J}\beta_{I)}=0\quad\then\quad
\6_{(J}\beta_{I)}=d_{IJ}=constant\label{m12}\eea
where we used that the $\6_I$ anticommute and that
$\6_{(J}\beta_{I)}$ has vanishing total degree. The
$d_{IJ}$ are thus symmetric
$\cG$-invariant constant tensors.
As $\beta_I$ has total degree 1, we conclude from
(\ref{m12}):
\beq \beta_I=h_I(\4\xi,\cT)+\4C^J(d_{JI}+h_{JI}(\cT)),\quad
h_{JI}=-h_{IJ}\ .\label{m13}\eeq
Applying $\6_K$ to (\ref{m11}) yields then
\beq \6_K\6_J\6_I\beta=\f JIL (h_{KL}+d_{KL})+\f KJL h_{IL}
+\f IKL h_{JL}\ .\label{m14}\eeq
Using (\ref{m7a}), (\ref{m11}), (\ref{m13}) and (\ref{m14})
it is now straightforward to determine first $\beta$ and
then $f$ by a calculation similar to the one that led to
(\ref{m10}). One finds
\beq G=4: \quad
 f=\4s\7h+\cH\7g+\4Q(\4s \7g)+d_{JI}\cF^I\cF^J
\label{m16}\eeq
where $\7g$ and $\7h$ depend only on the $\4\xi$ and $\cT$.
Differentiating (\ref{m16}) w.r.t. $\4Q$ yields
$\4s \7g=0$ which implies $\7g=\4s k(\cT)$ by
(\ref{c2}) and (\ref{r33}). Using
$\4s\cH=0$ and defining
$h=\7h-\cH k$, this finally results in
\beq G=4:\quad f=\4s h(\4\xi,\cT)+d_{JI}\cF^I\cF^J
\label{m17}\eeq
and completes the proof of (\ref{lem71}).

In the case $G=2$,
(\ref{lem72}) can be proved as follows. Assume
\[ a_{i_a}\cF^{i_a}=\4s h(\4\xi,\cT)\]
holds for some $h$.
Due to $\cF^{i_a}=\4s \4C^{i_a}$ for abelian $\cF$'s, this implies
\[ \4s\left[a_{i_a}\4C^{i_a}-h(\4\xi,\cT)\right]=0 \]
and thus, by (\ref{c2}) and (\ref{r33}),
\beq  a_{i_a}\4C^{i_a}=h(\4\xi,\cT)+\4s\, g(\cT)
\label{m18}\eeq
for some $g(\cT)$. However, as no $\4s\cT$ contains
a linear combination of the $\4C^{i_a}$ (with
constant coefficients), the
left and the right hand side of (\ref{m18}) must
vanish separately which implies indeed $a_{i_a}=0$.
Analogously one can treat the cases $G=3,4$
and complete the proof of (\ref{lem72}).

\mysection{Linearized weak supersymmetry cohomology}\label{global}

In this appendix we will compute the
weak cohomology of $\delta_{susy}$ in the space of
$\cG$-invariant local total forms $f(\4\xi,\7\cT)$ at total
degrees $\geq 4$
(cf. eq. (\ref{f11})). It will be shown that 
this cohomology vanishes
at all total degrees exceeding 4 and is at total degree 4
represented
by $\cG$-invariant local total forms
\beq
\7P =\7D_\da \7D^\da \, \Xi\,
\{ \cA(\5W,\5\lambda)+
D^\alpha D_\alpha \cB(\7\cT)\}+c.c.
\label{weak1}\eeq
with $\Xi$ as in (\ref{cp14}) and
$\7D_\da$ as in (\ref{cp15a}).
Furthermore we will derive the results for
total degree 3 presented in section \ref{step1}.

Let us therefore consider
\beq \delta_{susy}f(\4\xi,\7\cT)\sim 0,\quad
\delta_If=0,\qd
\tot(f)=G\geq 3.\label{weak2}\eeq
The problem will be analyzed along the lines
of \cite{glusy} to which I refer for details
(see section 6 and appendix A of \cite{glusy}).
I note however that the case $G=3$ was not treated in \cite{glusy}
and deserves therefore some special attention.

We decompose both $\delta_{susy}$ and $f$ according to the
degree in the $\4\xi^a$. To this end we introduce the counting operator
\beq
\4N=\4\xi^a\, \frac{\6}{\6\4\xi^a}\ .
\label{cp1}\eeq
$\delta_{susy}$ decomposes into three pieces with $\4N$-degrees
$1,0,-1$ respectively,
\beq\delta_{susy}=\delta_- +\delta_0+\delta_+\ ,
\quad [\4N,\delta_\pm]=\pm\delta_\pm\ ,\quad
[\4N,\delta_0]=0.\label{cp2}\eeq
These pieces are spelled out
in table E.1 where $b$ and $\5b$ are the operators
\beq b\, \7\cT=\4\xi^\alpha D_\alpha \7\cT\ ,\quad
\5b\, \7\cT=\4\xi^\da \5D_\da \7\cT\ ,\quad
b\,\4\xi^A=\5b\, \4\xi^A=0.
\label{cp9}\eeq
Here the linearized (weak)
$D_\alpha$-transformations given in table C.3 are to be used.
\[
\ba{c|c|c|c}
\cW            &  \delta_-\cW & \delta_0\cW & \delta_+\cW\\
\hline\rule{0em}{3ex}
\4\xi^{\da\alpha}   &  4i\4\xi^\alpha\4\xi^\da & 0 & 0 \\
\rule{0em}{3ex}
\4\xi^\alpha & 0 & 0 & 0 \\
\rule{0em}{3ex}
\4\xi^\da & 0 & 0 & 0 \\
\rule{0em}{3ex}
\7\cT     & 0 &
         (b+\5b)\, \7\cT
         & \4\xi^a D_a\7\cT  \\
\multicolumn{4}{c}{}\\
\multicolumn{4}{c}{\mbox{Table E.1: Decomposition of $\delta_{susy}$}}
\ea
\]
$\delta_{susy}f\sim 0$ decomposes into
\bea 0&= & \delta_-X_\up,\label{cp4}\\
0&\sim & \delta_-X_{\up+1}+\delta_0X_\up,\label{cp5}\\
0&\sim & \delta_-X_{p+1}+\delta_0X_p+\delta_+X_{p-1}\quad
\mbox{for $\up<p<\op$},\label{cp5a}\\
0&\sim &\delta_0X_\op+\delta_+X_{\op-1},\label{cp5b}\\
0&\sim &\delta_+X_\op\label{cp5c}
\eea
where $X_p$ is the part of $f(\4\xi,\cT)$ with degree $p$
in the $\4\xi^a$,
\beq f(\4\xi,\cT)=\sum_{p=\up}^\op X_p\ ,\quad
\4N X_p=pX_p\ .
\label{cp6}\eeq
Note that in (\ref{cp4}) we have used $=$ rather than $\sim$
as $\delta_-$ does not see the tensor fields at all.
Since $f(\4\xi,\7\cT)$ is defined only modulo weakly
$\delta_{susy}$-exact local total forms $\delta_{susy} h(\4\xi,\7\cT)$,
$X_\up$ can be assumed to represent
a nontrivial cohomology class of the
cohomology of $\delta_-$.
That cohomology has been determined in \cite{phd,dixon}
(see also \cite{glusy,sugra}).
The result is that a $\delta_-$-closed function depends,
up to $\delta_-$-exact pieces, on the $\4\xi^a$ only
via the quantities
\beq \th^\alpha= \4\xi_\da \4\xi^{\da\alpha}\ ,\quad
\5\th^\da=\4\xi^{\da\alpha}\xi_\alpha \ ,\qd
\Theta=\4\xi_\da \4\xi^{\da\alpha}\4\xi_\alpha\ .
                             \label{lem33}\eeq
Moreover the dependence on these quantities is very
restricted as one has
\beq \delta_-f(\4\xi^A)=0 \ \LRA \ f=P(\5\th^\da,\4\xi^\alpha)+
P'(\th^\alpha,\4\xi^\da)+a\, \Theta+\delta_-h(\4\xi^A)
\label{lem31}\eeq
where $a$ is constant. Furthermore no non-vanishing
function $P(\5\th^\da,\4\xi^\alpha)+
P'(\th^\alpha,\4\xi^\da)+a\, \Theta$ is $\delta_-$-exact,
\beq P(\4\th^\da,\4\xi^\alpha)+
P'(\4\th^\alpha,\4\xi^\da)+a\,\Theta=\delta_-h(\4\xi^A)
\ \LRA\ P=-P'=constant,\ a=0.
\label{lem32}\eeq
These results are very useful when analyzing
(\ref{cp4})--(\ref{cp5c}). First we use
(\ref{lem31}) to conclude that $X_\up$ 
can be assumed to be of the form
\bea
G\geq 4:& &
X_\up=P(\5\th^\da,\4\xi^\alpha,\7\cT)+
\5P(\th^\alpha,\4\xi^\da,\7\cT),\label{weak4}\\
G=3:& &
X_\up=P(\5\th^\da,\4\xi^\alpha,\7\cT)+
\5P(\th^\alpha,\4\xi^\da,\7\cT)+4\, \delta_\up^1\, \Theta\,
R(\7\cT)\label{weak5}\eea
where the Kronecker symbol $\delta_\up^1$ occurs as
$\Theta$ is linear in the $\4\xi^a$ and a factor 4 has been
introduced for later convenience. Furthermore we can
assume without loss of generality that $P$ and $\5P$ are
related by complex conjugation and that $R$ is real
(cf. second remark at the end of section \ref{reduce}).
Note that $\Theta$ has total degree 3 and can therefore occur
only for $G=3$ which complicates the analysis of this case
as compared to the other ones. Note also that
in fact we have $\up\in\{0,1,2\}$ as
the $\th$'s anticommute.

By inserting now (\ref{weak4}) resp. (\ref{weak5}) in (\ref{cp5})
we obtain, using (\ref{lem32})
\beq b\, P\sim 0\label{weak6}\eeq
with $b$ as in (\ref{cp9}). Furthermore we can assume without
loss of generality
\beq P\not\sim b\, Q\label{weak66}\eeq 
because otherwise
$P$ can be removed from $f$ by subtracting a suitable
$\delta_{susy}$-exact piece from $f$ without changing
the form of $X_\up$, i.e. without reintroducing a $\delta_-$-exact
piece in it (of course $P$, $\5P$ or $R$ may get
redefined).
Moreover $P$ is required to be $\cG$-invariant.
Using $P=\5\th^{\da_1}\cdots\5\th^{\da_\up}
\omega_{\da_1\cdots\da_\up}(\4\xi^\alpha,\7\cT)$
one concludes that $\omega_{\da_1\cdots\da_\up}$
is determined by the weak cohomology of $b$ on
$l_{\alpha\beta}$-invariant functions $f(\4\xi^\alpha,\7\cT)$ 
where $l_{\alpha\beta}$ generates Lorentz (resp.
$SL(2,C)$) transformations of undotted spinor indices
according to (\ref{ltrafo}). 

In order to compute the latter cohomology,
we need the QDS-structure
of the on-shell $D_\alpha$-representation proved in appendix \ref{onqds}.
Thanks to this structure we can directly adopt the analysis and
results of appendix A of \cite{glusy} to conclude
\beq bf(\4\xi^\alpha,\7\cT)\sim 0,\
l_{\alpha\beta}f=0
\ \LRA\ f\sim 
\cA(\5W,\5\lambda)+D^2 \cB(\7\cT)+bh(\4\xi^\alpha,\7\cT)
\label{lem41}\eeq
where $\cB$ and $h$ are
$l_{\alpha\beta}$-invariant and we used $D^2=D^\alpha D_\alpha$.
This result is the key to the solution of (\ref{weak2})
in the cases $G\geq 4$. Indeed, it implies in particular that
$\omega_{\da_1\cdots\da_\up}$ can be chosen so as not to depend
on $\4\xi^\alpha$ at all, and thus that the total degree of $P$
equals two times its degree in the $\5\th$'s. 
As the latter are Grassmann odd and linear in
the $\4\xi^a$, we
conclude immediately that
$P$ can be assumed to vanish in all cases $G\geq 3$ except for
the case $G=4$, $\up=2$ where (\ref{lem41}) yields
\beq G=4,\ \up=2:\quad P=-4i\, \5\th_\da\5\th^\da
[\cA(\5W,\5\lambda)+D^2 \cB(\7\cT)].
\label{cp11}\eeq
In particular this implies that each solution to
(\ref{weak2}) with $G>4$ is indeed trivial. Furthermore,
using (\ref{lem31}) again, it is easy to verify that
the equations (\ref{cp5a})--(\ref{cp5c}) do not
impose further obstructions in the case $G=4$ and lead to
the solutions (\ref{weak1}).
This completes the investigation of (\ref{weak2}) for $G\geq 4$.

We are thus left with the case $G=3$, $\up=1$ for which
(\ref{weak5}) reduces to
\beq G=3,\ \up=1:\quad X_1=4\Theta\, R(\7\cT).\label{weak9}\eeq
Here we used already that $P$ can be assumed to
vanish in this case by subtracting a trivial piece from $f$
and redefining $R$ suitably,
if necessary. Recall that so far we have only used
equations (\ref{cp4}) and (\ref{cp5}).
We now have to analyze the restrictions on the function $R(\7\cT)$
imposed by the remaining equations (\ref{cp5a})--(\ref{cp5c}).
To that end we need
the explicit form of $X_2$ corresponding to (\ref{weak9}).
It is obtained from (\ref{cp5}) and reads
\beq G=3,\ \up=1:\quad X_2=
-i(\5\th_\da\4\xi^{\da\alpha}D_\alpha
+\th^\alpha\4\xi_{\alpha\da}\5D^\da)\, R(\7\cT)
\label{weak10}\eeq
up to a $\delta_-$-exact piece which can be neglected
with no loss of generality. (\ref{cp5a}) now requires
\beq \delta_+X_1+\delta_0X_2+\delta_-X_3\sim 0\label{weak11}\eeq
for some $X_3$.
Elementary algebra with spinor indices yields straightforwardly
\beq \delta_0X_2=
i(-\th^\alpha\5\th^\da [D_\alpha,\5D_\da]
+\Theta\4\xi^{\da\alpha}\{D_\alpha,\5D_\da\}
+\s0 12\5\th\5\th\, D^2+\s0 12\th\th\, \5D^2)\,  R(\7\cT).
\label{weak11a}\eeq
Now, the first term
in (\ref{weak11a}), involving the commutator
$[D_\alpha,\5D_\da]$, is $\delta_-$-exact by
(\ref{lem31}) as it is $\delta_-$-closed and involves
both $\th$ and $\5\th$. The second term, involving the anticommutator
$\{D_\alpha,\5D_\da\}$, cancels exactly with
$\delta_+X_1$
in (\ref{weak11}) due to $\{D_\alpha,\5D_\da\}R\sim -2iD_{\alpha\da}R$.
The remaining two terms in
(\ref{weak11a}), involving
$D^2R$ and $\5D^2R$ respectively, are $\delta_-$-closed but
not $\delta_-$-exact, see (\ref{lem31}) and (\ref{lem32}). Hence,
(\ref{weak11}) requires
\beq D^2R(\7\cT)\sim 0.\label{weak13}\eeq
Moreover, (\ref{weak11}) now determines $X_3$ unambigously as
$X_4$ vanishes in the case $G=3$. One finds
\beq X_3=\s0 1{12}\, \4\xi^{\da\beta}
\4\xi_{\beta\dbe}\4\xi^{\dbe\alpha}
[D_\alpha,\5D_\da]\,  R(\7\cT).\label{weak12a}\eeq
$X_1+X_2+X_3$ yields now indeed (\ref{f12a}).

\end{document}